\documentclass[a4paper,11pt]{article}
\pdfoutput=1 % if your are submitting a pdflatex (i.e. if you have
\usepackage{jheparxiv} % for details on the use of the package, please
\usepackage[T1]{fontenc} % if needed

\usepackage{tensor}
\usepackage{amsmath}
\usepackage{amssymb}
\usepackage{mathrsfs}
\usepackage{graphicx}
\usepackage{stmaryrd}
\usepackage{twistor}
\usepackage{commath}
\usepackage{xcolor}
\usepackage{mathtools}

\newcommand{\bigma}{\boldsymbol{\sigma}}

\newcommand{\msf}[1]{\mathsf{#1}}

\newcommand{\Res}[1]{\underset{#1}{\text{Res}}\,}

\newcommand{\bmu}{\boldsymbol{\mu}}

\newcommand{\ba}{\mathbf{a}}

\newcommand{\m}{\mathrm{m}}

\newcommand{\nbar}{\bar{\nabla}}

\title{Infrared structures of scattering on self-dual radiative backgrounds}

\author{Tim Adamo,}
\author{Wei Bu}
\author{\& Bin Zhu}

\affiliation{School of Mathematics and Maxwell Institute for Mathematical Sciences \\
University of Edinburgh, EH9 3FD, U.K.}

\emailAdd{t.adamo@ed.ac.uk}
\emailAdd{w.bu@sms.ed.ac.uk} 
\emailAdd{bzhu@exseed.ed.ac.uk}

\begin{document} 

\abstract{The scattering of gluons and gravitons in trivial backgrounds is endowed with many surprising infrared features which have interesting conformal interpretations on the two-dimensional celestial sphere. However, the fate of these structures in more general asymptotically flat backgrounds is far from clear. In this paper, we consider holomorphic infrared structures in the presence of non-perturbative, self-dual background gauge and gravitational fields which are determined by freely specified radiative data. We make use of explicit formulae for tree-level gluon and graviton scattering in these self-dual radiative backgrounds, as well as chiral twistor sigma model descriptions of the classical dynamics. Remarkably, we find that the leading holomorphic part of tree-level collinear splitting functions -- or celestial OPEs -- and infinite-dimensional chiral soft algebras are undeformed by the background. We also compute all-order holomorphic celestial OPEs in the MHV sectors of gauge theory and gravity.}

\maketitle
\flushbottom

\section{Introduction}

Celestial holography aims to describe physics in four dimensional asymptotically flat spacetime as a hologram on the two-dimensional celestial sphere, which can be encoded in a putative \emph{celestial conformal field theory} (CCFT)~\cite{Strominger:2017zoo, Raclariu:2021zjz,Pasterski:2021rjz,Pasterski:2021raf,McLoughlin:2022ljp}. The bottom-up approach to celestial holography seeks to exploit underlying 2d conformal structures in bulk scattering amplitudes (the natural holographic observables in asymptotically flat settings) to deduce fundamental properties of any candidate CCFT. This has been particularly fruitful in the infrared (IR) regime. Collinear limits of massless particles have the structure of an operator product expansion (OPE) on the celestial sphere, and in the conformal primary basis~\cite{Pasterski:2016qvg,Pasterski:2017kqt,Pasterski:2017ylz} this enables one to identify OPE coefficients between gluon or graviton insertions on the celestial sphere~\cite{Fan:2019emx,Pate:2019lpp,Banerjee:2020kaa,Fotopoulos:2020bqj,Banerjee:2020zlg,Banerjee:2020vnt}. Soft expansions in the conformal primary basis~\cite{Donnay:2018neh,Adamo:2019ipt,Puhm:2019zbl,Guevara:2019ypd} have led to the discovery of infinite-dimensional chiral algebras in perturbative gauge theory and gravity~\cite{Guevara:2021abz,Strominger:2021lvk,Himwich:2021dau,Jiang:2021ovh}, associated with the underlying integrability of the self-dual sector~\cite{Adamo:2021lrv}. Indeed, it is clear that the 2d CFT perspective on 4d massless scattering has been extremely powerful, regardless of any holographic interpretation. 

It is natural to wonder at the extent to which any of the beautiful structures discovered in this context persist in more general asymptotically flat settings. To date, the vast majority of explorations of celestial holography have been confined to perturbative Yang-Mills and general relativity (and their EFT extensions) in trvial backgrounds: flat gauge fields and Minkowski spacetime. There have been some investigations in this direction recently, which can be roughly divided into two streams. The first explores scattering in some sourced asymptotically flat background (e.g., coupled dilaton profiles, shockwaves, black holes) using the source to regularize the behaviour of celestial amplitudes relative to flat backgrounds~\cite{Fan:2022vbz,Casali:2022fro,Fan:2022kpp,deGioia:2022fcn,Gonzo:2022tjm,Stieberger:2022zyk, Melton:2022fsf,Banerjee:2023rni,Taylor:2023bzj,Stieberger:2023fju}. 

The second explores the fate of celestial OPEs and infinite-dimensional chiral algebras in the setting where non-trivial \emph{self-dual} (SD) asymptotically flat background fields are introduced. It seems that these structures are quite often deformed by SD background gauge and gravitational fields: non-commutativity in SD gravity on Minkowski space~\cite{Bu:2022iak,Monteiro:2022lwm,Monteiro:2022xwq}, monopoles~\cite{Garner:2023izn}, Burns space~\cite{Costello:2022jpg,Costello:2023hmi} and the Eguchi-Hansen gravitational instanton~\cite{Bittleston:2023bzp} all lead to deformations of collinear splitting functions and/or the chiral algebra associated to the SD sector. However, these case-by-case studies do little to paint a unifying, guiding picture about how, and under what circumstances, chiral CCFT structures are deformed in the presence of asymptotically flat background fields. Furthermore, in all of these examples there are no concrete formulae for the actual background field scattering amplitudes (or at least not beyond 3-points), further hampering the ability to make general statements. 

\medskip

In this paper, we explore the structure of holomorphic collinear splitting functions and infinite-dimensional chiral algebras in Yang-Mills theory and general relativity on a broad class of complex asymptotically flat gauge and gravitational backgrounds: \emph{self-dual radiative backgrounds}. These are SD solutions to the vacuum equations of motion which are determined by free radiative data on null infinity. In other words, SD radiative backgrounds include unconstrained \emph{functional} degrees of freedom: they form an open set in the space of asymptotically flat, SD solutions to the field equations.

From the traditional perspective of background field theory, studying scattering amplitudes or their IR features on SD radiative backgrounds looks hopeless due to the freely specified radiative data and the lack of any other simplifying symmetry assumptions. This also ensures that many modern amplitudes methods are useless in these backgrounds, as their scattering amplitudes will not be rational functions of the asymptotic kinematics. However, SD radiative backgrounds have an elegant description in terms of twistor theory, which enables \emph{all-multiplicity} computations of tree-level gluon and graviton scattering amplitudes~\cite{Adamo:2020syc,Adamo:2020yzi,Adamo:2022mev} through the use of classical, chiral 2d CFTs called `twistor sigma models'~\cite{Adamo:2021bej}\footnote{Indeed, the twistor description is so powerful that it also enables the computation of partially off-shell observables in SD radiative backgrounds, such as (supersymmetric) form factors of local operators in Yang-Mills theory~\cite{Bogna:2023bbd}.}.

We compute the holomorphic part\footnote{Throughout, the `holomorphic part' of a collinear splitting function or celestial OPE means the part which is singular in the kinematic invariant $\la i\,j\ra$, rather than $[i\,j]$, written in terms of spinor helicity variables for the external momenta of collinear particles $i$ and $j$.} of the leading, tree-level collinear splitting functions of gluons in SD radiative gauge fields and gravitons in SD radiative spacetimes. We perform these calculations using both the explicit, all-multiplicity formulae for the tree-level MHV scattering amplitudes, and the 2d twistor sigma models. We also use the twistor sigma models to compute the chiral algebras associated to the SD sectors of gauge theory and gravity in these backgrounds.

Remarkably, we find that the holomorphic collinear splitting functions, celestial OPEs and chiral algebras are \emph{undeformed} from those in a trivial background. On a technical level, this seems to be related to the fact that the holomorphic structures on twistor space associated with SD radiative backgrounds can be given global representations. Our results suggest that in celestial holography for the SD sector, holomorphic IR structures are only deformed by the introduction of sources or `large' data in the bulk -- a feature shared by all of the examples studied previously in the literature. Furthermore, our results have interesting implications in the MHV sector, where we show that the holomorphic celestial OPE can be computed explicitly to \emph{all orders}, including all finite contributions. Compared to the flat background cases~\cite{Adamo:2022wjo,Ren:2023trv}, regular terms in the OPEs are dependent on the background data. 

\medskip

The paper is structured as follows. We begin in section \ref{sec:review} with a review of SD radiative backgrounds, their twistor theory and scattering amplitudes. In section \ref{sec:holOPEgauge}, we compute the leading, holomorphic part of the collinear splitting function,
or celestial OPE, between external gluons scattering on a SD radiative gauge field background in two ways: from the explicit scattering amplitudes formulae and from the OPEs in the twistor sigma model. In section \ref{sec:holOPEgravity}, we examine the gravity amplitudes in similar ways, finding some new subtleties in the mixed helicity case. In section \ref{sec5}, we study the infinite-dimensional chiral algebras in SD radiative backgrounds. Section \ref{sec6} leverages these results to compute the all-order holomorphic OPEs in the MHV sector for SD radiative backgrounds. Section \ref{sec:dis} concludes, and Appendix~\ref{appA} provides details of a conjectural generalization of the graviton S-matrix on SD radiative spacetimes which manifests collinear splitting for all helicity configurations.

%%%%%%%%%%%%%%%%%%%%%
%%%%%%%%%%%%%%%%%%%%%

\section{Self-dual radiative backgrounds and twistor sigma models} \label{sec:review}

In this section, we review the description of scattering in self-dual (SD) radiative backgrounds in gauge theory and gravity through twistor theory. Broadly speaking, these backgrounds are asymptotically flat, complex (in Lorentzian signature), chiral solutions of the vacuum Yang-Mills or Einstein equations which are specified by a single complex function of three variables, specifying the radiative data at infinity. This function constitutes the free characteristic data of the non-linear field equations, and as such is totally unconstrained, aside from basic regularity assumptions and having the appropriate spin and conformal weights.

The functional freedom inherent in SD radiative solutions makes them extremely challenging to study as scattering backgrounds for gauge theory or gravity with the traditional background field expansion. However, their description in terms of twistor theory is remarkably elegant and can be operationalized to enable perturbative calculations using \emph{twistor strings}~\cite{Witten:2003nn,Berkovits:2004hg,Berkovits:2004jj,Mason:2007zv,Skinner:2013xp} or \emph{twistor sigma models}~\cite{Adamo:2021bej}: chiral 2d CFTs governing holomorphic structures in twistor space. This has enabled derivations of the all-multiplicity MHV amplitudes for gluons and gravitons on SD radiative gauge fields and spacetimes, respectively~\cite{Adamo:2020syc}, and well-motivated conjectures for the full tree-level S-matrices in both cases~\cite{Adamo:2020yzi,Adamo:2022mev}. These formulae, and the twistor sigma models underpinning them, are the basic tools that we use to investigate collinear splitting and soft algebras in SD radiative backgrounds.

%%%%%%%%%%%%%%%%%%%%%

\subsection{Yang-Mills}

The future null conformal boundary, $\scri^+$, of four-dimensional Minkowski spacetime $\M$ has topology $\R\times S^2$ and is usually described in terms of coordinates $(u,\zeta,\bar{\zeta})$, with $u$ the retarded Bondi time and $(\zeta,\bar{\zeta})$ stereographic coordinates on the sphere. Despite their ubiquity, these coordinates break manifest Lorentz invariance (as the stereographic coordinates are not global on the sphere), and for our purposes it is preferable to instead use a manifestly Lorentz-invariant description of $\scri^+$. This is provided by a homogeneous description of $\scri^+$ as the total space of a line bundle over the sphere~\cite{Eastwood:1982,Sparling:1990,Adamo:2014yya,Geyer:2014lca,Adamo:2015fwa,Adamo:2021dfg}.

In particular, we adopt a homogeneous set of coordinates
\be\label{scr1}
(u,\lambda_{\alpha},\bar{\lambda}_{\dot\alpha})\sim(|b|^2u,\,b\lambda_{\alpha},\,\bar{b}\bar{\lambda}_{\dot\alpha})\, \quad \forall b\in\C^*\,,
\ee
where $(\lambda_{\alpha},\bar{\lambda}_{\dot\alpha})$ are the homogeneous coordinates on $\P^1$. This presents $\scri^+$ as the total space of the line bundle $\cO_{\R}(1,1)\rightarrow\P^1$, sections of which are real-valued homogeneous functions $f(\lambda,\bar{\lambda})$ obeying $f(b\lambda,\bar{b}\bar{\lambda})=|b|^2\,f(\lambda,\bar{\lambda})$, and the degenerate conformal metric on $\scri^+$ is
\be\label{scri2}
\d\hat{s}^2_{\scri^+}=0\times\d u^2+\D\lambda\,\D\bar{\lambda}\,, \qquad \D\lambda:=\la\lambda\,\d\lambda\ra\,,\quad \D\bar{\lambda}:=[\bar{\lambda}\,\d\bar{\lambda}]\,.
\ee
The conformal boundary can be (minimally) complexified by allowing $u$ to take complex values, in which case $\scri_{\C}$ is the total space of the complex line bundle $\cO(1,1)\to\P^1$. 

One of the many advantages of this description is that spin- and conformal-weighted functions are given by sections of the line bundles $\cO(p,q)\to\scri_\C$, with spin-weight corresponding to $s=\frac{p-q}{2}$ and conformal-weight to $w=\frac{p+q}{2}$. For instance, the radiative data of a complex, asymptotically flat solution of the Yang-Mills equations on $\M$ is given by a pair of adjoint-valued functions $\cA^{0}(u,\lambda,\bar{\lambda})$, $\tilde{\cA}^{0}(u,\lambda,\bar\lambda)$ valued in $\cO(-2,0)$ and $\cO(0,-2)$, controlling the anti-self-dual and self-dual degrees of freedom, respectively~\cite{vanderBurg:1969,Exton:1969im,Ashtekar:1981bq,Strominger:2013lka,Barnich:2013sxa}. Besides some rudimentary regularity assumptions (e.g., that the total radiated energy is finite), this radiative data is freely specified. A \emph{radiative} gauge field is an asymptotically flat Yang-Mills field which is uniquely determined by this radiative data~\cite{vanderBurg:1969,Newman:1978ze}, and a \emph{self-dual radiative} gauge field is a complex radiative Yang-Mills field which is purely self-dual~\cite{Goldberg:1979wt,Newman:1980fr,Adamo:2020yzi}. Equivalently, a SD radiative gauge field is uniquely characterised by the radiative data $\tilde{\cA}^{0}$, with $\cA^0=0$. The class of SD radiative gauge fields includes, for example, self-dual plane waves but \emph{not} self-dual gauge fields with sources or `large' data, such as the self-dual dyon or BPST instanton.

From the traditional perspective of background field theory, the task of computing gluon scattering amplitudes on a SD radiative background gauge field appears hopeless: the background is specified by a free function of three variables, $\tilde{\cA}^0(u,\lambda,\bar{\lambda})$, with no other symmetry assumptions. However, SD radiative gauge fields have a remarkably simple descriptions in terms of twistor theory. The twistor space of $\M$ is the open subset of $\P^3$
\be\label{PT}
\PT=\left.\left\{Z^{A}=(\mu^{\dot\alpha},\lambda_{\alpha})\in\P^3\,\right|\,\lambda_{\alpha}\neq 0\right\}\,,
\ee
where $Z^{A}=(\mu^{\dot\alpha},\lambda_{\alpha})$ are homogeneous coordinates on $\P^3$. The relation between $\PT$ and $\M$ is captured by the incidence relations
\be\label{inc}
\mu^{\dot\alpha}=x^{\alpha\dot\alpha}\,\lambda_{\alpha}\,,
\ee
implying that a point $x\in\M$ corresponds to a holomorphic, linearly embedded $X\cong\P^1$ in $\PT$. There is a natural projection from twistor space to $\scri_\C$ given by
\be\label{scriproj}
p:\PT\to\scri_\C\,, \qquad (\mu^{\dot\alpha},\lambda_\alpha)\mapsto (u=[\mu\,\bar{\lambda}],\,\lambda_{\alpha},\,\bar{\lambda}_{\dot\alpha})\,,
\ee
which is clearly compatible with the homogeneous description \eqref{scr1}.

The Ward correspondence~\cite{Ward:1977ta} gives an equivalence between SD gauge fields on spacetime and holomorphic vector bundles $E\to\PT$ on twistor space satisfying some straightforward technical assumptions. The holomorphic structure of the vector bundle is encoded in a partial connection $\bar{D}:\Omega^{p,q}(\PT,E)\to\Omega^{p,q+1}(\PT,E)$ subject to $\bar{D}^2=0$. For SD radiative gauge fields, this partial connection has a global realization in terms of the radiative data~\cite{Sparling:1990,Newman:1978ze,Newman:1980fr}:
\be\label{SDradgf}
\bar{D}=\dbar+p^{*}\tilde{\cA}^0\,\D\bar{\lambda}\,,
\ee
which automatically obeys $\bar{D}^2=0$ as a consequence of $\D\bar{\lambda}\wedge\D\bar{\lambda}=0$. 

This makes it straightforward to couple twistor string theory to SD radiative background gauge fields. In particular, the gauge theoretic degrees of freedom are encoded in twistor string theory by a 2d worldsheet current algebra on the Riemann sphere. For concreteness, we take the gauge group to be SU$(N)$ and realise the current algebra as a complex free fermion system coupled to the partial connection encoding the SD radiative background:
\be\label{curralg}
S_{C}=\frac{1}{2\pi}\int_{\Sigma}\rho_{i}\,\bar{D}|_{\Sigma}\bar{\rho}^{i}\,,
\ee
where $\Sigma$ is a closed Riemann surface and  $\rho_i$ is a fermion valued in $K^{1/2}_{\Sigma}$ (i.e., of conformal weight $\frac{1}{2}$) and the fundamental representation of the gauge group and $\bar{\rho}^{i}$ is a fermion valued in $K^{1/2}_{\Sigma}$ and the anti-fundamental representation of the gauge group (so $i,j=1,\ldots,N$). The partial connection $\bar{D}|_{\Sigma}$ here is understood in terms of pullback via a holomorphic map $Z^{A}:\Sigma\to\PT$.

While the OPE of the fermion system remains free,
\be\label{ffope}
\rho_i(\sigma)\,\bar{\rho}^{j}(\sigma')\sim\frac{\delta_{i}^{j}}{\sigma-\sigma'}\,,
\ee
in terms of a holomorphic affine coordinates on $\Sigma$, correlations functions are sensitive to the background gauge field. Indeed, at genus zero $\Sigma\cong\P^1$ and we can use homogeneous coordinates $\sigma^{\ba}=(\sigma^0,\sigma^1)$ on the Riemann sphere, for which the propagator is
\be\label{ffcorr}
\left\la\rho_{i}(\sigma)\,\bar{\rho}^{j}(\sigma')\right\ra=\frac{\sqrt{\D\sigma\,\D\sigma'}}{(\sigma\,\sigma')}\,\msf{H}_{i}{}^{k}(\sigma)\,\msf{H}^{-1\, j}{}_{k}(\sigma')\,,
\ee
where $(\sigma\,\sigma'):=\sigma^{\ba}\,\sigma'_{\ba}$ is the M\"obius-invariant inner product on $\P^1$ and $\D\sigma:=(\sigma\,\d\sigma)$. Here, the key ingredient is $\msf{H}$, the holomorphic frame which trivializes the partial connection \eqref{SDradgf} on rational curves; it is defined implicitly by:
\be\label{holframe}
\msf{H}^{-1}\,\bar{D}|_{\Sigma}\msf{H}=\dbar|_{\Sigma}\,.
\ee
The existence of such a holomorphic trivialization is generic for the vector bundles $E\to\PT$ arising in the Ward correspondence~\cite{Sparling:1990}, and always exists for SD radiative gauge fields. 

The adjoint-valued worldsheet current is constructed as usual
\be\label{wsc}
j^{\msf{a}}(\sigma):=(\msf{T}^{\msf{a}})_{i}{}^{j}\,\rho_{i}(\sigma)\,\bar{\rho}^{j}(\sigma)\,,
\ee
where $\msf{T}^{\msf{a}}$ are the generators, $\msf{a}=1,\ldots,N^2-1$ and normal-ordering is implicit. The OPE structure of the current algebra is the standard one dictated by conformal invariance:
\be\label{jjOPE}
j^{\msf{a}}(\sigma)\,j^{\msf{b}}(\sigma')\sim \frac{k\,\delta^{\msf{ab}}}{(\sigma-\sigma')^2}+\frac{f^{\msf{abc}}\,j^{\msf{c}}(\sigma')}{\sigma-\sigma'}\,,
\ee
where $k$ is the level. However, as a consequence of \eqref{ffcorr}, correlation functions of these objects are modified by the holomorphic frames. Indeed, at genus zero, the effective propagator between worldsheet currents becomes
\begin{multline}\label{jjprop}
\left\la j^{\msf{a}}(\sigma)\,j^{\msf{b}}(\sigma')\right\ra= \\ \frac{\D\sigma\,\D\sigma'}{(\sigma\,\sigma')^2}\left[\tr\left(\msf{H}^{-1}(\sigma)\,\msf{T}^{\msf{a}}\,\msf{H}(\sigma)\,\msf{H}^{-1}(\sigma')\,\msf{T}^{\msf{b}}\,\msf{H}(\sigma')\right)+\tr\left(\msf{H}(\sigma)\,\msf{T}^{\msf{a}}\,\msf{H}^{-1}(\sigma)\,\msf{H}(\sigma')\,\msf{T}^{\msf{b}}\,\msf{H}^{-1}(\sigma')\right)\right]\\
+\frac{\sqrt{\D\sigma\,\D\sigma'}}{(\sigma\,\sigma')}\left(\rho(\sigma)\,\msf{T}^{\msf{a}}\msf{H}(\sigma)\,\msf{H}^{-1}(\sigma')\,\msf{T}^{\msf{b}}\,\bar{\rho}(\sigma')-\rho(\sigma')\,\msf{T}^{\msf{b}}\,\msf{H}^{-1}(\sigma')\,\msf{H}(\sigma)\,\msf{T}^{\msf{a}}\,\bar{\rho}(\sigma)\right)\,,
\end{multline}
with the terms proportional to the double pole arising due to multiple Wick contractions among the constituents of the worldsheet currents. 

\medskip

Gluon scattering amplitudes in the SD radiative background can then be computed by calculating correlation functions in this background-coupled worldsheet current algebra. External positive and negative helicity gluons are encoded by vertex operators
\be\label{gluVO}
\cU^{\msf{a}}_{\pm}=\int_{\Sigma}j^{\msf{a}}(\sigma)\,a_{\pm}(Z(\sigma))\,,
\ee
where $a_{\pm}\in H^{0,1}(\PT,\cO(\pm2-2))$ represent the free gluon wavefunction on twistor space via the Penrose transform~\cite{Penrose:1969ae,Eastwood:1981jy}. In the negative helicity case, \eqref{gluVO} does not really make sense as the integrand does not have balanced homogeneity on $\Sigma\cong\P^1$. To make $\cU^{\msf{a}}_{-}$ well-defined, one must include in the integrand some object $\msf{O}(Z(\sigma))$ valued in $\cO(4)$ on $\PT$; this is usually accomplished with $\cN=4$ target space supersymmetry. However, for our purposes it suffices to leave $\msf{O}$ implicit in the definition of $a_{-}$ as it plays no dynamical role in the 2d CFT and is always evaluated on zero modes in any correlation function. 

For momentum eigenstates, where the external gluon has asymptotic on-shell momentum $k_{\alpha\dot\alpha}=\kappa_{\alpha}\,\tilde{\kappa}_{\dot\alpha}$, these twistor wavefunctions are given by~\cite{Adamo:2011pv}
\be\label{glumomeig}
a_{\pm}(Z)=\int_{\C^*}\frac{\d t}{t^{\pm2-1}}\,\bar{\delta}^{2}(\kappa_{\alpha}-t\,\lambda_\alpha)\,\e^{\im\,t\,[\mu\,\tilde{\kappa}]}\,,
\ee
with the scale integral over $t$ ensuring the wavefunction has the correct homogeneity and
\be\label{holdelt}
\bar{\delta}^{2}(\kappa_{\alpha}-t\,\lambda_\alpha):=\frac{1}{(2\pi\im)^2}\bigwedge_{\alpha=0,1}\dbar\left(\frac{1}{\kappa_{\alpha}-t\,\lambda_\alpha}\right)\,,
\ee
the holomorphic delta function. Twistor wavefunctions for the conformal primary basis are easily obtained from these by taking 
\be\label{Mellinmom}
\kappa_{\alpha}=z_{\alpha}=(1,z)\,, \qquad \tilde{\kappa}_{\dot\alpha}=\varepsilon\,\omega\,\bar{z}_{\dot\alpha}=\varepsilon\,\omega\,(1,\bar{z})\,,
\ee
and Mellin transforming with respect to $\omega$, for $\varepsilon=\pm1$ labelling outgoing/incoming states\footnote{This slightly non-standard chiral assignment of the frequency $\omega$ is well-adapted to the holomorphic celestial OPEs, and has the consequence of giving little group weight to coordinates on the celestial sphere~\cite{Adamo:2022wjo}.}. As a result of the chiral assignment of $\omega$, the Mellin transform for a helicity $\pm1$ gluon must include an additional power of $\omega^{\mp1}$ in the measure to obtain the correct little group scaling~\cite{Adamo:2022wjo}.

Now, the computation of gluon scattering proceeds in one of two equivalent ways: either by setting the level of the current algebra to zero ($k\to0$) or by taking only the semi-classical correlation function of the worldsheet currents. This is necessary to decouple unwanted (indeed, non-unitary) gravitational modes from the correlator~\cite{Berkovits:2004jj,Adamo:2018hzd}, and in practical terms amounts to discarding double-pole contributions to the propagator \eqref{jjprop} and keeping only single Wick contractions between any two currents. 

The resulting expression for the colour-ordered tree-level N$^{d-1}$MHV gluon amplitude in the SD radiative background is~\cite{Adamo:2020yzi}:
\be\label{gluamp1}
\cA_{n,d}=\int\frac{\d^{4(d+1)}U}{\mathrm{vol}\,\GL(2,\C)}\,\tr\!\left(\prod_{i=1}^{n}\msf{H}_{i}^{-1}\,\msf{T}^{\msf{a}_i}\,\msf{H}_i\right)|\tilde{\mathtt{g}}|^4\,\prod_{j=1}^{n}\frac{a_{\pm,j}\,\D\sigma_j}{(i\,i+1)}\,,
\ee
Here, $U_{(\alpha_1\cdots\alpha_d)}^{A}$ are the moduli of the degree $d$ holomorphic map $Z^A:\P^1\to\PT$
\be\label{holmap}
Z^{A}(\sigma)=U^{A}_{(\ba_1\cdots\ba_d)}\,\sigma^{(\ba_1}\cdots\sigma^{\ba_d)}\equiv U^{A}_{\ba(d)}\,\sigma^{\ba(d)}\,,
\ee
$\msf{H}_i:=\msf{H}(U,\sigma_i)$, $a_{\pm,i}:=a_{\pm,i}(Z(\sigma_i))$, $\tilde{\mathtt{g}}\subset\{1,\ldots,n\}$ denotes the set of $d+1$ negative helicity gluons and
\be\label{glvdm}
|\tilde{\mathtt{g}}|=\prod_{\substack{i,j\in\tilde{\mathtt{g}} \\ i<j}}(i\,j)\,,
\ee
is the Vandermonde determinant associated to the negative helicity gluons. 

For all $d>1$, the formula \eqref{gluamp1} is conjectural, in the sense that it is not derived from the perturbative expansion of Yang-Mills theory in the background field formalism. Nevertheless, they pass a multitude of non-trivial consistency checks, including the perturbative limit (where the SD radiative background becomes a single positive helicity gluon) and background gauge invariance~\cite{Adamo:2020yzi}. In the MHV sector, where $d=1$, the formulae \emph{can} be derived directly from the Yang-Mills action on $\M$, and further simplification is possible as the GL$(2,\C)$ redundancy in the integral measure over the map moduli can be used to identify the homogeneous coordinate $\sigma_\alpha$ with $\lambda_\alpha$, the twistor coordinates on the celestial sphere. The MHV amplitude, with gluons $a,b\in\{1,\ldots,n\}$ negative helicity, becomes~\cite{Adamo:2020syc,Adamo:2020yzi}
\be\label{MHVglunab}
\cA_{n,1}=\frac{\la a\,b\ra^{4}}{\la1\,2\ra\,\la2\,3\ra\cdots\la (n-1)\,n\ra\,\la n\,1\ra}\,\int\d^{4}x\,\tr\!\left(\prod_{i=1}^{n}\msf{H}^{-1}(x,\kappa_i)\,\msf{T}^{\msf{a}_i}\,\msf{H}(x,\kappa_i)\right)\,\e^{\im\,\sum_{j=1}^{n}k_j\cdot x}\,,
\ee
for a general SD radiative background gauge field. Here, the formula is presented for momentum eigenstates, with all Riemann sphere integrals having been performed against the holomorphic delta functions appearing in \eqref{glumomeig}.

%%%%%%%%%%%%%%%%%%%%%

\subsection{Gravity}

The homogeneous description of $\scri_\C$ extends to any complexified, asymptotically flat spacetime, so it can also be used to describe SD radiative spacetimes. These are complex, vacuum, asymptotically flat spacetimes with self-dual Weyl tensor which are uniquely determined by their radiative data on $\scri_\C$. The radiative information in any asymptotically flat spacetime is controlled by the asymptotic shear of the outgoing null hypersurfaces of constant $u$~\cite{Sachs:1961zz,Jordan:1961}; these are encoded by the Newman-Penrose scalars $\bigma^{0}(u,\lambda,\bar{\lambda})$ and $\tilde{\bigma}^0(u,\lambda,\bar{\lambda})$, corresponding to the ASD and SD degrees of freedom and valued in $\cO(-3,1)$ and $\cO(1,-3)$, respectively~\cite{Newman:1961qr}. For Lorentzian real metrics, these are related by complex conjugation, but in the complexified setting or with Euclidean or $(2,2)$-signature reality conditions, they are independent.

A radiative spacetime is an asymptotically simple spacetime which solves the vacuum Einstein equations and is uniquely determined by this radiative data, with vanishing Bondi mass aspect as $u\to\infty$~\cite{Friedrich:1986rb}. A SD radiative spacetime is then a complex radiative spacetime with data $\tilde{\bigma}^0$ and $\bigma^0=0$. These complex spacetimes (sometimes called $\cH$-spaces) are tautologically holographic: the full SD metric can be reconstructed from the asymptotic radiative data~\cite{Newman:1976gc,Hansen:1978jz}. In particular, they are a special case of the twistorial non-linear graviton construction~\cite{Penrose:1976js}. This gives a one-to-one correspondence between SD vacuum Einstein metrics and integrable complex deformations $\CPT$ of the flat twistor space $\PT$ which preserve a 4-parameter family of holomorphic rational curves along with a fibration $\CPT\to\P^1$ and the trivial Poisson structure on the fibres. 

For SD radiative spacetimes, this can be manifested by writing the complex structure on $\CPT$ explicitly in terms of the radiative data~\cite{Eastwood:1982}. This complex structure is captured by a Dolbeault operator $\nbar:\Omega^{p,q}(\CPT)\to\Omega^{p,q+1}(\CPT)$ obeying $\nbar^2=0$, and for SD radiative spacetimes this operators has a global realization as
\be\label{SDradmet}
\nbar=\dbar+p^{*}\tilde{\bigma}^0\,\D\bar{\lambda}\,\bar{\lambda}_{\dot\alpha}\,\frac{\partial}{\partial\mu^{\dot\alpha}}\,,
\ee
where $\dbar$ is the standard complex structure on $\P^3$ with holomorphic homogeneous coordinates $(\mu^{\dot\alpha},\lambda_{\alpha})$. This automatically obeys $\nbar^2=0$, again as a consequence of $\D\bar{\lambda}\wedge\D\bar{\lambda}=0$. 

As in the gauge theory setup, the computation of graviton scattering in a SD radiative spacetime can be operationalized through computing semiclassical correlation functions in a 2d CFT. In this case, the CFT is a chiral twistor sigma model~\cite{Adamo:2021bej} describing holomorphic rational curves in $\CPT$ with respect to the complex structure \eqref{SDradmet}. Such a degree $d$ holomorphic curve can be described by
\be\label{holcurv}
Z^{A}(\sigma)=\left(\mu^{\dot\alpha}(\sigma)=\msf{F}^{\dot\alpha}(U,\sigma),\,\lambda_{\alpha}(\sigma)\right)\,,
\ee
where $\lambda_{\alpha}(\sigma)=U_{\alpha\,\ba(d)}\sigma^{\ba(d)}$ and $\msf{F}^{\dot\alpha}$ is homogeneous of degree $d$ on the curve and obeys
\be\label{holcurveeq}
\dbar|_{\Sigma}\msf{F}^{\dot\alpha}(U,\sigma)=\bar{\lambda}^{\dot\alpha}(\sigma)\,\tilde{\bigma}^{0}\,\D\bar{\lambda}(\sigma)\,.
\ee
Determining this $\msf{F}^{\dot\alpha}$ explicitly is the gravitational equivalent of determining the holomorphic frame $\msf{H}$ in the gauge theory construction. 

At degree $d$, the computation of the N$^{d-1}$MHV amplitude corresponds to calculating tree-level correlation functions controlled by the chiral CFT~\cite{Adamo:2022mev}
\be\label{tsigm}
S=\int_{\Sigma}\frac{\D\sigma}{\mathfrak{s}^2_{\tilde{\mathtt{h}}}(\sigma)}\,m^{\dot\alpha}\,\left(\epsilon_{\dot\beta\dot\alpha}\,\dbar|_{\Sigma} -\bar{\lambda}_{\dot\alpha}\,\bar{\lambda}_{\dot\beta}\,N([\msf{F}\,\bar{\lambda}],\lambda,\bar{\lambda})\,\D\bar{\lambda}\right)m^{\dot\beta}\,,
\ee
where the dynamical field $m^{\dot\alpha}$ is a section of $\cO(d)\to\Sigma\cong\P^1$, 
\be\label{sdef}
\mathfrak{s}_{\tilde{\mathtt{h}}}(\sigma):=\prod_{k\in\tilde{\mathtt{h}}}(\sigma\,k)\,,
\ee
is a section of $\cO(d+1)\to\Sigma\cong\P^1$ defined by the set of $d+1$ points $\tilde{\mathtt{h}}\subset\{1,\ldots,n\}$ corresponding to the negative helicity gravitons, and $N(u,\lambda,\bar{\lambda})$ is the self-dual \emph{news function} of the background:
\be\label{news}
N:=-\frac{\partial\tilde{\bigma}^0}{\partial u}\,, \qquad N^{(p)}:=\frac{\partial^p N}{\partial u^p}\,.
\ee 
This quantity is invariant under BMS supertranslations and encodes the (complex) radiation flux at infinity~\cite{Sachs:1962zza,Bondi:1962px,Sachs:1962wk}.

The correlation functions of interest involve vertex operators for positive helicity external gravitons
\be\label{gravVO}
\cU_{+}=\int_{\Sigma}\frac{\D\sigma}{\mathfrak{s}^2_{\tilde{\mathtt{h}}}(\sigma)}\,h_+(\msf{F}+m,\sigma)\,,
\ee
where $h_+\in H^{0,1}_{\nbar}(\CPT,\cO(2))$ is the twistor wavefunction~\cite{Hitchin:1980hp}, and vertex operators for the SD radiative background itself:
\be\label{backVO}
\cU^{(p)}=\frac{2}{p!}\int_{\Sigma}\frac{\D\sigma}{\mathfrak{s}^2_{\tilde{\mathtt{h}}}(\sigma)}\wedge\D\bar{\lambda}\,[m\,\bar{\lambda}]^{p}\,N^{(p-2)}([\msf{F}\,\bar{\lambda}],\lambda,\bar{\lambda})\,,
\ee
with $N^{(p)}:=\frac{\partial^p N}{\partial u^p}$. These background vertex operators encode tail effects in the scattering amplitude~\cite{Bonnor:1959,Bonnor:1966,Thorne:1980ru,Blanchet:1987wq}, where an external graviton interacts with the background which can subsequently self-interact. 

The CFT \eqref{tsigm} is free, so the short distance OPE is trivial
\be\label{tsigOPE}
m^{\dot\alpha}(\sigma)\,m^{\dot\beta}(\sigma')\sim\frac{\epsilon^{\dot\alpha\dot\beta}}{\sigma-\sigma'}\,,
\ee
but correlation functions are sensitive to the SD radiative background through the propagator
\be\label{mprop}
\left\la m^{\dot\alpha}(\sigma)\,m^{\dot\beta}(\sigma')\right\ra=\frac{H^{\dot\alpha}{}_{\dot\gamma}(U,\sigma)\,H^{\dot\beta\dot\gamma}(U,\sigma')}{(\sigma\,\sigma')}\,\mathfrak{s}_{\tilde{\texttt{h}}}(\sigma)\,\mathfrak{s}_{\tilde{\texttt{h}}}(\sigma')\,,
\ee
where $H^{\dot\alpha}{}_{\dot\beta}$ provides a holomorphic frame for the dotted spinor bundles over the degree $d$ holomorphic curves \eqref{holcurveeq} in $\CPT$. It can be defined by
\be\label{dotframe1}
\frac{\partial\msf{F}^{\dot\alpha}}{\partial U^{\dot\beta}_{\ba(d)}}(U,\sigma)=H^{\dot\alpha}{}_{\dot\beta}(U,\sigma)\,\sigma^{\ba(d)}\,,
\ee
and obeys
\be\label{dotframe2}
H_{\dot\gamma\dot\alpha}(U,\sigma)\,H^{\dot\gamma}{}_{\dot\beta}(U,\sigma)=\epsilon_{\dot\alpha\dot\beta}\,, \qquad \dbar|_{\Sigma}H^{\dot\alpha}{}_{\dot\beta}=\bar{\lambda}^{\dot\alpha}\,\bar{\lambda}_{\dot\gamma}\,H^{\dot\gamma}{}_{\dot\beta}\,N\,\D\bar{\lambda}\,,
\ee
in terms of the news function.

\medskip

The calculation of graviton scattering amplitudes proceeds by perturbatively expanding a natural family of generating functionals for the N$^{d-1}$MHV amplitudes through computing connected, tree correlators of these vertex operators in the 2d CFT of \eqref{tsigm}. The resulting tree-level N$^{d-1}$MHV amplitude for graviton scattering on a SD radiative spacetime is~\cite{Adamo:2022mev}:
\begin{multline}\label{gramp1}
\cM_{n,d}=\sum_{t=0}^{n-d-3}\sum_{p_1,\ldots,p_t>2}\int\frac{\d^{4(d+1)}U}{\mathrm{vol\,\GL(2,\C)}}\,|\tilde{\mathtt{h}}|^8\,\mathrm{det}'\!\left(\HH^{\vee}\right)\left(\prod_{\m=1}^{t}\,\frac{\partial^{p_\m}}{\partial\varepsilon_\m^{p_\m}}\right)\mathrm{det}^{\prime}(\cH)\Big|_{\varepsilon=0}\\
\times \prod_{i=1}^{n}h_{\pm,i}(Z(\sigma_i))\,\prod_{\m=1}^{t}\D\bar{\lambda}(\sigma_m)\,N^{(p_\m-2)}(\sigma_\m)\,.
\end{multline}
Here, the sums range over the number of background interactions and their multiplicity, while the object $\mathrm{det}'(\HH^{\vee})$ is the resultant of the $\lambda_\alpha(\sigma)$ components of the map~\cite{Skinner:2013xp,Cachazo:2013zc}\footnote{The reason for the suggestive notation $\mathrm{det}'(\HH^{\vee})\equiv R(\lambda_{\alpha})$ is that in Minkowski spacetime, $\cH\to\HH$, the Hodges matrix, and $\mathrm{det}'(\HH)$ is exchanged with $\mathrm{det}'(\HH^{\vee})$ under parity~\cite{Bullimore:2012cn,Cachazo:2012pz}. In a SD background, this is no longer the case, as the background is parity-violating.}: it is independent of all marked points on $\P^1$, vanishes if $\lambda_{\alpha}(\sigma)=0$ (ensuring that $\cM_{n,d}$ has support only when the holomorphic curve lies in $\CPT\subset\P^3$) and respects the structure of boundary divisors in the space of holomorphic maps. The $h_{\pm,i}\in H^{0,1}_{\nbar}(\CPT,\cO(\pm4-2))$ are the twistor wavefunctions for the external gravitons; for momentum eigenstates, these are:
\be\label{gravmomeig}
h_{\pm}(Z)=\int\frac{\d t}{t^{\pm4-1}}\,\bar{\delta}^{2}(\kappa_{\alpha}-t\,\lambda_{\alpha})\,\e^{\im\,t\,[\msf{F}\,\tilde{\kappa}]}\,,
\ee
which are in the cohomology of $\nbar$ defined by \eqref{SDradmet} as a consequence of $\D\bar{\lambda}\wedge\D\bar{\lambda}=0$.

The final ingredient in this formula is the reduced determinant $\mathrm{det}'(\cH)$ of the $(n+t-d-1)\times(n+t-d-1)$ matrix $\cH$, which has the block decomposition
\be\label{cHblock}
\cH=\left(\begin{array}{c c}
          \HH & \mathfrak{h} \\
          \mathfrak{h}^{\mathrm{T}} & \T
          \end{array}\right)\,,
\ee
The $(n-d-1)\times(n-d-1)$ block has indices corresponding to each positive helicity external graviton $i,j\notin\tilde{\mathtt{h}}$:
\be\label{HHblock}
\HH_{ij}=-t_i\,t_j\,\frac{[\![i\,j]\!]}{(i\,j)}\,\sqrt{\D\sigma_i\,\D\sigma_j}, \quad i\neq j\,,
\ee
\begin{equation*}
\HH_{ii}=t_{i}\,\D\sigma_i\sum_{j\neq i}\,\frac{t_j\,[\![i\,j]\!]}{(i\,j)}\prod_{l\in\tilde{\mathtt{h}}}\,\frac{(l\,j)}{(l\,i)}-\im\,t_i\,\D\sigma_i\sum_{\m=1}^{t}\,\frac{\varepsilon_\m\,[\![i\,\bar{\lambda}(\sigma_\m)]\!]}{(i\,\m)}\prod_{l\in\tilde{\mathtt{h}}}\,\frac{(l\,\m)}{(l\,i)}\,,
\end{equation*}
where 
\be\label{dressedmom}
[\![i\,j]\!]=\tilde{K}^{\dot\alpha}_{i}\,\tilde{K}_{j\,\dot\alpha}:=\tilde{\kappa}_{i\,\dot\beta}\,\tilde{\kappa}_{j\,\dot\gamma}\,H^{\dot\beta\dot\alpha}(U,\sigma_i)\,H^{\dot\gamma}{}_{\dot\alpha}(U,\sigma_j)\,,
\ee
with $\tilde{K}_{i}^{\dot\alpha}$ denoting the background-dressed dotted momentum spinor. The $(n-d-1)\times t$ block $\mathfrak{h}$ has entries
\be\label{hblock}
\mathfrak{h}_{i\m}=\im\,t_i\,\varepsilon_{\m}\,\frac{[\![i\,\bar{\lambda}(\sigma_\m)]\!]}{(i\,\m)}\,\sqrt{\D\sigma_i\,\D\sigma_{\m}}\,,
\ee
and the final $t\times t$ block $\T$ is given by
\be\label{Tblock}
\T_{\m\mathrm{n}}=\varepsilon_{\m}\,\varepsilon_{\mathrm{n}}\,\frac{[\![\bar{\lambda}(\sigma_\m)\,\bar{\lambda}(\sigma_{\mathrm{n}})]\!]}{(\m\,\mathrm{n})}\,\sqrt{\D\sigma_\m\,\D\sigma_{\mathrm{n}}}\,, \quad \m\neq\mathrm{n}\,,
\ee
\begin{equation*}
\T_{\m\m}=-\im\,\varepsilon_\m\,\D\sigma_\m\sum_{i=1}^{n}\,\frac{t_i\,[\![\bar{\lambda}(\sigma_\m)\,i]\!]}{(\m\,i)}\,\prod_{l\in\tilde{\mathtt{h}}}\,\frac{(l\,i)}{(l\,\m)}-\varepsilon_\m\,\D\sigma_\m\sum_{\mathrm{n}\neq\m}\,\frac{\varepsilon_{\mathrm{n}}\,[\![\bar{\lambda}(\sigma_\m)\,\bar{\lambda}(\sigma_{\mathrm{n}})]\!]}{(\m\,\mathrm{n})}\,\prod_{l\in\tilde{\mathtt{h}}}\,\frac{(l\,\mathrm{n})}{(l\,\m)}\,.
\end{equation*}
The reduced determinant itself is defined by
\be\label{reddet}
\mathrm{det}^{\prime}(\cH):=\frac{|\cH^{i}_{i}|}{|\tilde{\mathtt{h}}\cup\{i\}|^2}\,,
\ee
where $|\cH^{i}_{i}|$ is the minor of $\cH$ obtained by removing the row and column corresponding to any $i\neq\tilde{\mathtt{h}}$, with the denominator corresponding to a Vandermonde determinant, defined analogously to \eqref{glvdm}.

As in gauge theory, for $d>1$ these amplitudes are conjectural, insofar as there is no derivation of them following directly from the Einstein-Hilbert action, although they pass multiple consistency tests. However, for the MHV sector ($d=1$), the formula can be derived directly from Plebanski's chiral formulation of general relativity~\cite{Plebanski:1977zz} is further simplified by fixing the GL$(2,\C)$ redundancy of the moduli measure to identify $\sigma_{\ba}$ with $\lambda_{\alpha}$ and performing all $\P^1$ integrals against delta functions. Taking gravtions $1,2$ to have negative helicity, the tree-level MHV graviton amplitude is given by
\begin{multline}\label{MHVgrav}
\cM_{n,1}=\la1\,2\ra^8\,\sum_{t=0}^{n-4}\sum_{p_1,\ldots,p_t>2}\int\d^{4}x\,\sqrt{g}\,\left(\prod_{\m=1}^{t}\,\frac{\partial^{p_\m}}{\partial\varepsilon_\m^{p_\m}}\right)\mathrm{det}^{\prime}(\cH)\Big|_{\varepsilon=0}\\
\times \exp\!\left(\im\sum_{i=1}^{n}\msf{F}^{\dot\alpha}(x,\kappa_i)\,\tilde{\kappa}_{i\,\dot\alpha}\right)\,\prod_{\m=1}^{t}\D\lambda_\m\wedge\D\bar{\lambda}_\m\,N^{(p_\m-2)}(\lambda_\m)\,,
\end{multline}
as the resultant $\mathrm{det}'(\HH^{\vee})$ of a degree one map is the identity. Here, all the $\P^1$ integrals associated with external gravitons have been performed against holomorphic delta functions, simplifying the entries of $\cH$ to:
\be\label{HHblockMHV}
\HH_{ij}=-\frac{[\![i\,j]\!]}{\la i\,j\ra}\,, \quad i\neq j\,,
\ee
\begin{equation*}
\HH_{ii}=\sum_{j\neq i}\,\frac{[\![i\,j]\!]}{\la i\,j\ra}\,\frac{\la1\,j\ra\,\la 2\,j\ra}{\la1\,i\ra\,\la2\,i\ra}-\im\,\sum_{\m=1}^{t}\,\frac{\varepsilon_\m\,[\![i\,\bar{\lambda}_\m]\!]}{\la\,i\,\lambda_\m\ra}\,\frac{\la1\,\lambda_\m\ra\,\la2\,\lambda_\m\ra}{\la1\,i\ra\,\la2\,i\ra}\,,
\end{equation*}
\be\label{hblockMHV}
\mathfrak{h}_{i\m}=\im\,\varepsilon_{\m}\,\frac{[\![i\,\bar{\lambda}_\m]\!]}{\la i\,\lambda_\m\ra}\,,
\ee
and
\be\label{TblockMHV}
\T_{\m\mathrm{n}}=\varepsilon_{\m}\,\varepsilon_{\mathrm{n}}\,\frac{[\![\bar{\lambda}_\m\,\bar{\lambda}_{\mathrm{n}}]\!]}{\la\lambda_\m\,\lambda_\mathrm{n}\ra}\,, \quad \m\neq\mathrm{n}\,,
\ee
\begin{equation*}
\T_{\m\m}=-\im\,\varepsilon_\m\sum_{i=1}^{n}\,\frac{[\![\bar{\lambda}_\m\,i]\!]}{\la\lambda_\m\,i\ra}\,\frac{\la1\,i\ra\,\la2\,i\ra}{\la1\,\lambda_\m\ra\,\la2\,\lambda_\m\ra}-\varepsilon_\m\,\sum_{\mathrm{n}\neq\m}\,\frac{\varepsilon_{\mathrm{n}}\,[\![\bar{\lambda}_\m\,\bar{\lambda}_{\mathrm{n}}]\!]}{\la\lambda_\m\,\lambda_\mathrm{n}\ra}\,\frac{\la1\,\lambda_\mathrm{n}\ra\,\la2\,\lambda_{\mathrm{n}}\ra}{\la1\,\lambda_\m\ra\,\la2\,\lambda_\m\ra}\,.
\end{equation*}
Here $(\lambda_\m,\bar{\lambda}_\m)$ are the homogeneous coordinates on the $t$ copies of the celestial sphere associated with each explicit insertion of the background news function.

%%%%%%%%%%%%%%%%%%%%%
%%%%%%%%%%%%%%%%%%%%%

\section{Gauge theory: leading holomorphic splitting} \label{sec:holOPEgauge}

In this section, we compute the leading, holomorphic part of the collinear splitting function, or celestial OPE, between external gluons scattering on a SD radiative gauge field background. There are two ways of doing this calculation: directly from the explicit scattering amplitude formulae on these backgrounds and using the vertex operators of the semi-classical twistor sigma model. Both procedures give the same result: the leading holomorphic collinear splitting function is undeformed by the background, for all helicity configurations.

%%%%%%%%%%%%%%%%%%%%%

\subsection{From the MHV amplitude}\label{subsec:gauge_MHV}

We begin by considering the collinear limit of tree-level gluon scattering amplitudes in a SD radiative gauge field. To do this, we parametrize the collinear regime between any two external particles $i,j$ as
\be\label{collim}
k_i +k_j=P+\varepsilon^2\,q\,,
\ee
where $P^{\alpha\dot\alpha}=\kappa_P^{\alpha}\,\tilde{\kappa}_P^{\dot\alpha}$ is the on-shell collinear momentum, $q^{\alpha\dot\alpha}=\xi^{\alpha}\,\tilde{\xi}^{\dot\alpha}$ is an arbitrary null reference vector, and the parameter $\varepsilon$ controls the collinear limit. As $k_i\cdot k_j=\varepsilon^2\,P\cdot q$, the holomorphic collinear limit corresponds to $\la i\,j\ra\sim\varepsilon\to0$. In terms of the parametrization \eqref{Mellinmom}, this is equivalent to $z_i-z_j:=z_{ij}\to0$ on the (complexified) celestial sphere, making the holomorphic collinear limit equivalent to a holomorphic celestial OPE limit~\cite{Fan:2019emx}.

In this holomorphic collinear limit, 
\be\label{collms}
\kappa_i^{\alpha}=\frac{\la\xi\,i\ra}{\la\xi\,P\ra}\,\kappa_P^{\alpha}+O(\varepsilon)\,, \qquad \kappa_j^{\alpha}=\frac{\la\xi\,j\ra}{\la\xi\,P\ra}\,\kappa_P^{\alpha}+O(\varepsilon)\,.
\ee
Applying these limits to MHV helicity configurations is sufficient to capture the holomorphic part of collinear limits in gauge theory or gravity.

\medskip

Now, let us begin with the formula \eqref{MHVglunab} for colour-ordered, tree-level MHV gluon scattering in a general, non-abelian SD radiative background gauge field. Without loss of generality, we can consider the holomorphic celestial OPE limit where positive helicity external gluons $i, j$ become collinear. Non-trivial holomorphic celestial OPEs will then be proportional to $\la i\,j\ra^{-1}=z_{ij}^{-1}$. Fixing all other external gluons to be in the same colour-ordering, only the $(\cdots i\,j\cdots)$ and $(\cdots j\,i\cdots)$ orderings will have the desired celestial OPE singularities.

From \eqref{MHVglunab}, this gives\footnote{We abuse notation throughout by denoting the holomorphic collinear limit of an amplitude by $\lim_{\varepsilon\to0}\cA_n$, which is clearly un-defined; by this we mean: `the leading singular part of $\cA_n$ as $\varepsilon\to0$.' More precisely, this is captured by $\lim_{\varepsilon\to0}\varepsilon\,\cA_n$, but we trust the reader will understand our simplified notation.}
\begin{multline}\label{glpp1}
\lim_{\varepsilon\to0}\left[\cA_{n,1}(\ldots,i^+,j^+,\ldots) +\cA_{n,1}(\ldots,j^+,i^+,\ldots)\right] \\
=\frac{1}{\la i\,j\ra}\,\frac{\la\xi\,P\ra^2}{\la\xi\,i\ra\,\la\xi\,j\ra}\,\frac{\la a\,b\ra^{4}}{\la1\,2\ra\cdots\la i-1 P\ra\,\la P\,j+1\ra\cdots\la n\,1\ra} \int\d^{4}x\,\exp\!\left[\im\left(P\cdot x+\sum_{l\neq i,j}k_l\cdot x\right)\right] \\ \left[\tr\left(\cdots\msf{H}^{-1}\!\left(\frac{\la\xi\,i\ra}{\la\xi\,P\ra}\kappa_P\right)\msf{T}^{\msf{a}_i}\msf{H}\!\left(\frac{\la\xi\,i\ra}{\la\xi\,P\ra}\kappa_P\right)\msf{H}^{-1}\!\left(\frac{\la\xi\,j\ra}{\la\xi\,P\ra}\kappa_P\right)\msf{T}^{\msf{a}_j}\msf{H}\!\left(\frac{\la\xi\,j\ra}{\la\xi\,P\ra}\kappa_P\right)\cdots\right)\right. \\
-\left.\tr\left(\cdots\msf{H}^{-1}\!\left(\frac{\la\xi\,j\ra}{\la\xi\,P\ra}\kappa_P\right)\msf{T}^{\msf{a}_j}\msf{H}\!\left(\frac{\la\xi\,j\ra}{\la\xi\,P\ra}\kappa_P\right)\msf{H}^{-1}\!\left(\frac{\la\xi\,i\ra}{\la\xi\,P\ra}\kappa_P\right)\msf{T}^{\msf{a}_i}\msf{H}\!\left(\frac{\la\xi\,i\ra}{\la\xi\,P\ra}\kappa_P\right)\cdots\right)\right],
\end{multline}
where the spacetime position dependence of the holomorphic frames has been suppressed for brevity. The holomorphic frames $\msf{H}(x,\lambda)$ are matrix valued functions which are homogeneous of degree zero in $\lambda$, meaning that
\be\label{framescale}
\msf{H}\!\left(x,\,\frac{\la\xi\,i\ra}{\la\xi\,P\ra}\kappa_P\right)=\msf{H}(x,\kappa_P)\,,
\ee
and similarly for $j$. 

This simplifies \eqref{glpp1} to
\begin{multline}\label{glpp2}
\frac{1}{\la i\,j\ra}\,\frac{\la\xi\,P\ra^2}{\la\xi\,i\ra\,\la\xi\,j\ra}\,\frac{\la a\,b\ra^{4}}{\la1\,2\ra\cdots\la i-1 P\ra\,\la P\,j+1\ra\cdots\la n\,1\ra} \int\d^{4}x\,\exp\!\left[\im\left(P\cdot x+\sum_{l\neq i,j}k_l\cdot x\right)\right] \\ \times\,\tr\left(\cdots\msf{H}^{-1}(x,\kappa_P)\,[\msf{T}^{\msf{a}_i},\,\msf{T}^{\msf{a}_j}]\,\msf{H}(x,\kappa_P)\cdots\right)\,,
\end{multline}
which in turn gives the splitting formula
\begin{multline}\label{glpp3}
\lim_{\varepsilon\to0}\left[\cA_{n,1}(\ldots,i^+,j^+,\ldots) +\cA_{n,1}(\ldots,j^+,i^+,\ldots)\right] \\
=\mathrm{Split}(i_1^+,j_1^+\to P_1^+)\,\cA_{n-1,1}(\cdots,P^+,\cdots)\,,
\end{multline}
where the holomorphic collinear splitting function
\be\label{glppsplit}
\mathrm{Split}(i_1^+,j_1^+\to P_1^+)=\frac{f^{\msf{a}_i\msf{a}_j\msf{c}}}{\la i\,j\ra}\,\frac{\la\xi\,P\ra^2}{\la\xi\,i\ra\,\la\xi\,j\ra}\,.
\ee
This is the well-known tree-level collinear splitting function of positive helicity gluons in a trivial perturbative background~\cite{Altarelli:1977zs,Mangano:1990by,Birthwright:2005ak}, so the SD radiative background has no effect on the splitting function itself. Note that different choices of the spinor $\xi^{\alpha}$ do not alter the splitting function in the strict collinear limit, where $\la i\,j\ra\to0$.

A nearly identical calculation also describes the mixed helicity case, where a positive helicity gluon $i$ becomes collinear with a negative helicity external gluon $a$. Following the same steps as before, one soon arrives at
\begin{multline}\label{glpn1}
\lim_{\varepsilon\to0}\left[\cA_{n,1}(\ldots,i^+,a^-,\ldots) +\cA_{n,1}(\ldots,a^-,i^+,\ldots)\right] \\
=\mathrm{Split}(i_1^+,a_1^-\to P_1^-)\,\cA_{n-1,1}(\cdots,P^-,\cdots)\,,
\end{multline}
with holomorphic collinear splitting function
\be\label{glpnsplit}
\mathrm{Split}(i_1^+,a_1^-\to P_1^-)=\frac{f^{\msf{a}_i\msf{a}_a\msf{c}}}{\la a\,i\ra}\,\frac{\la\xi\,a\ra^3}{\la\xi\,i\ra\,\la\xi\,P\ra^2}\,,
\ee
again matching the known result for tree-level gluon scattering in a trivial perturbative background~\cite{Altarelli:1977zs,Mangano:1990by,Birthwright:2005ak}.

\medskip

Of course, it follows immediately from \eqref{glppsplit} that the celestial OPE coefficient obtained by Mellin transform will also be the same as that in a trivial background. Nevertheless, it is instructive to see how this emerges; for simplicity, we take both collinear gluons to be outgoing (i.e., $\varepsilon_i=\varepsilon_j=+1$) with positive helicity, and use the parametrization \eqref{Mellinmom}. Performing the same calculation as above, but for the Mellin-transformed celestial amplitude, one need only keep track of the transforms for the two collinear particles.

Straightforward manipulations lead to an expression of the form
\be\label{glppOPE1}
\frac{f^{\msf{a}_i\msf{a}_j\msf{c}}}{z_{ij}}\int\limits_{\M\times\R_+^2}\!\!\d^{4}x\,\d\omega_i\,\omega_i^{\Delta_i-2}\,\d\omega_j\,\omega_j^{\Delta_j-2}\,\exp\!\left[\im\,(\omega_i+\omega_j)\left(\bar{z}_{j}^{\dot\alpha}+\frac{\bar{z}_{ij}^{\dot\alpha}}{1+\frac{\omega_j}{\omega_i}}\right)x_{\alpha\dot\alpha}\,z_{j}^{\alpha}\right]\,,
\ee
where the celestial amplitude is implicitly evaluated in the holomorphic OPE limit $z_{ij}\to0$ and all factors which do not depend on the frequencies of the collinear particles have been ignored. After performing the sequential rescalings $\omega_i\to\omega_i\omega_j$ followed by $\omega_j\to\omega_j (1+\omega_i)^{-1}$, the exponential can be Taylor expanded in $\bar{z}_{ij}$ to obtain:
\be\label{glppOPE2}
\frac{f^{\msf{a}_i\msf{a}_j\msf{c}}}{z_{ij}}\sum_{m=0}^{\infty}\,\,\int\limits_{\M\times\R_+^2}\!\!\d^{4}x\,\frac{\d\omega_i\,\omega_i^{\Delta_i+m-2}}{(1+\omega_i)^{\Delta_i+\Delta_j+m-2}}\,\d\omega_j\,\omega_j^{\Delta_i+\Delta_j-3}\,\frac{\bar{z}_{ij}^m}{m!}\,\dbar_j^m\,\e^{\im x_{\alpha\dot\alpha} z_i^{\alpha}\bar{z}_j^{\dot\alpha}}\,,
\ee
where $\dbar_j:=\frac{\partial}{\partial\bar{z}_j}$. By performing the integral over $\omega_i$ explicitly and recalling the definition of the conformal primary external wavefunctions, we obtain the leading, singular holomorphic OPE coefficient, including the full tower of regular anti-holomorphic descendants:
\be\label{glppOPE3}
C(\Delta_i^{+,\msf{a}_i},\Delta_j^{+,\msf{a}_j};(\Delta_i+\Delta_j-1)^{\msf{c}}|\bar{z}_{ij})=f^{\msf{a}_i\msf{a}_j\msf{c}}\sum_{m=0}^{\infty}B(\Delta_i+m-1,\Delta_j-1)\,\frac{\bar{z}_{ij}^m}{m!}\,\dbar_j^m\,,
\ee
where $B(x,y)$ is the Euler Beta function. As expected, this matches the result for the singular holomorphic celestial OPE coefficient in a trivial perturbative background~\cite{Fan:2019emx,Pate:2019lpp,Guevara:2021abz}. 

%%%%%%%%%%%%%%%%%%%%%

\subsection{From the twistor sigma model}\label{subsection_YM_TSM_calculation}

The same results for the collinear splitting functions can be obtained directly from the semiclassical worldsheet OPE between gluon vertex operators \eqref{gluVO} in the twistor sigma model \eqref{curralg}. As in a flat background~\cite{Adamo:2021zpw}, this computation makes no reference to the other external states in the scattering process, and as such makes sense for generic $d$. Rather than performing the calculation directly in the OPE limit on the CFT worldsheet $\Sigma\cong\P^1$, we simply use the classical limit of contractions between the worldsheet currents. In the background-coupled free fermion realization \eqref{curralg}, this is given by the simple pole contribution from \eqref{jjprop}:
\begin{multline}\label{scjj}
j^{\msf{a}}(\sigma)\,j^{\msf{b}}(\sigma')\Big|_{\text{class.}} \\
= \frac{\sqrt{\D\sigma\,\D\sigma'}}{(\sigma\,\sigma')}\left(\rho(\sigma)\,\msf{T}^{\msf{a}}\msf{H}(\sigma)\,\msf{H}^{-1}(\sigma')\,\msf{T}^{\msf{b}}\,\bar{\rho}(\sigma')-\rho(\sigma')\,\msf{T}^{\msf{b}}\,\msf{H}^{-1}(\sigma')\,\msf{H}(\sigma)\,\msf{T}^{\msf{a}}\,\bar{\rho}(\sigma)\right)\,,
\end{multline}
where $\msf{H}$ is the holomorphic frame associated to the Ward bundle $E\to\PT$ of the SD radiative background.

Let $\cU^{\msf{a}}_{+\,i}$, $\cU^{\msf{b}}_{+\,j}$ be vertex operators representing the collinear external gluons. Using \eqref{scjj}, one finds
\begin{multline}\label{glwsOPE1}
\cU^{\msf{a}}_{+\,i}\,\cU^{\msf{b}}_{+\,j}\Big|_{\text{class.}}= \int\frac{\sqrt{\D\sigma_i\,\D\sigma_j}}{(i\,j)}\left(\rho_i\,\msf{T}^{\msf{a}}\msf{H}_i\,\msf{H}^{-1}_j\,\msf{T}^{\msf{b}}\,\bar{\rho}_j-\rho_j\,\msf{T}^{\msf{b}}\,\msf{H}^{-1}_j\,\msf{H}_i\,\msf{T}^{\msf{a}}\,\bar{\rho}_i\right)\,\frac{\d t_i}{t_i}\,\frac{\d t_j}{t_j} \\
\bar{\delta}^{2}(\kappa_i-t_i\,\lambda(\sigma_i)) \,\bar{\delta}^{2}(\kappa_j-t_j\,\lambda(\sigma_j))\,\exp\!\left[\im\left(t_i\,[\mu(\sigma_i)\,i]+t_j\,[\mu(\sigma_j)\,j]\right)\right]\,,
\end{multline}
where we have abbreviated $\rho_i\equiv\rho(\sigma_i)$, $\bar{\rho}_i\equiv\bar{\rho}(\sigma_i)$, $\msf{H}_i\equiv\msf{H}(U,\sigma_i)$ and similarly for this objects evaluated at $\sigma_j$. Now, the scale integral in $t_i$ can be performed against one of the holomorphic delta functions in the integrand to leave
\begin{multline}\label{glwsOPE2}
-\int\frac{\sqrt{\D\sigma_i\,\D\sigma_j}}{(i\,j)}\left(\rho_i\,\msf{T}^{\msf{a}}\msf{H}_i\,\msf{H}^{-1}_j\,\msf{T}^{\msf{b}}\,\bar{\rho}_j-\rho_j\,\msf{T}^{\msf{b}}\,\msf{H}^{-1}_j\,\msf{H}_i\,\msf{T}^{\msf{a}}\,\bar{\rho}_i\right)\,\frac{\d t_j}{t_j}\,\frac{\la\xi\,\lambda(\sigma_i)\ra}{\la\xi\,i\ra} \\
\bar{\delta}(\la i\,\lambda(\sigma_i)\ra)\,
\bar{\delta}^{2}(\kappa_j-t_j\,\lambda(\sigma_j))\,\exp\!\left[\im\left(\frac{[\mu(\sigma_i)\,i]\,\la\xi\,i\ra}{\la\xi\,\lambda(\sigma_i)\ra}+t_j\,[\mu(\sigma_j)\,j]\right)\right]\,,
\end{multline}
with $\bar{\delta}(\la i\,\lambda(\sigma_i)\ra)$ a holomorphic delta function in $\sigma_i$.

At this point, we observe that this holomorphic delta function can be written as
\be\label{glholdel}
\bar{\delta}(\la i\,\lambda(\sigma_i)\ra)=\frac{1}{2\pi\im}\,\bar{D}_i\left(\frac{1}{\la i\,\lambda(\sigma_i)\ra}\right)\,,
\ee
for $\bar{D}_i$ the pullback of the partial connection \eqref{SDradgf} to $\Sigma$ acting at $\sigma_i$. This is true because the part of $\bar{D}_i$ encoding the background acts trivially on the argument of the holomorphic delta function, which is not coupled to the Ward bundle. Since $\bar{D}$ is integrable, we can now integrate by parts in \eqref{glwsOPE2}, moving $\bar{D}_i$ onto all other $\sigma_i$-dependent factors of the integrand. As a consequence of \eqref{holframe}, $\bar{D}_i\msf{H}_i=0$ and the integration-by-parts picks out the simple pole in $(i\,j)$, leaving:
\begin{multline}\label{glwsOPE3}
\int\sqrt{\D\sigma_i\,\D\sigma_j}\,\bar{\delta}(i\,j)\left(\rho_i\,\msf{T}^{\msf{a}}\msf{H}_i\,\msf{H}^{-1}_j\,\msf{T}^{\msf{b}}\,\bar{\rho}_j-\rho_j\,\msf{T}^{\msf{b}}\,\msf{H}^{-1}_j\,\msf{H}_i\,\msf{T}^{\msf{a}}\,\bar{\rho}_i\right)\,\frac{\d t_j}{t_j} \\
\frac{\la\xi\,\lambda(\sigma_i)\ra}{\la\xi\,i\ra\,\la i\,\lambda(\sigma_i)\ra} \,
\bar{\delta}^{2}(\kappa_j-t_j\,\lambda(\sigma_j))\,\exp\!\left[\im\left(\frac{[\mu(\sigma_i)\,i]\,\la\xi\,i\ra}{\la\xi\,\lambda(\sigma_i)\ra}+t_j\,[\mu(\sigma_j)\,j]\right)\right]\,.
\end{multline}
The integral in $\sigma_i$ can now be performed trivially against the holomorphic delta function in the first line, setting $\sigma_i=\sigma_j$ everywhere in the remaining integrand:
\begin{multline}\label{glwsOPE4}
f^{\msf{abc}}\,\frac{\la\xi\,j\ra}{\la i\,j\ra\,\la\xi\,i\ra}\int j^{\msf{c}}(\sigma_j)\,\frac{\d t_j}{t_j}\,\bar{\delta}^{2}(\kappa_j-t_j\,\lambda(\sigma_j)) \\
\times\,\exp\!\left[\frac{\im\,t_j}{\la\xi\,j\ra}\Big([\mu(\sigma_j)\,i]\,\la\xi\,i\ra+[\mu(\sigma_j)\,j]\,\la\xi\,j\ra\Big)\right]\,,   
\end{multline}
where the integrand has been evaluated on the support of the remaining holomorphic delta functions.

Using the collinear limit \eqref{collim}, \eqref{collms} and reparametrizing $t_j\to t_j \frac{\la\xi\,j\ra}{\la\xi\,P\ra}$, it is now straightforward to see that \eqref{glwsOPE4} is equivalent to
\be\label{glwsOPE5}
\cU^{\msf{a}}_{+\,i}\,\cU^{\msf{b}}_{+\,j}\Big|_{\text{class.}}=\frac{f^{\msf{abc}}}{\la i\,j\ra}\,\frac{\la\xi\,P\ra^2}{\la\xi\,i\ra\,\la\xi\,j\ra}\,\cU^{\msf{c}}_{+\,P}+O(\varepsilon)\,.
\ee
As expected, this matches \eqref{glppsplit}, producing exactly the same holomorphic collinear splitting function, or celestial OPE, as in a trivial background. A similar calculation, differing only in the homogeneity of the scaling parameter $t_j$, shows that   
\be\label{glwsOPE6}
\cU^{\msf{a}}_{+\,i}\,\cU^{\msf{b}}_{-\,j}\Big|_{\text{class.}}=\frac{f^{\msf{abc}}}{\la i\,j\ra}\,\frac{\la\xi\,j\ra^3}{\la\xi\,i\ra\,\la\xi\,P\ra^2}\,\cU^{\msf{c}}_{-\,P}+O(\varepsilon)\,,
\ee
matching \eqref{glpnsplit}. In other words, the classical 2d CFT of the twistor sigma model dynamically produces the holomorphic collinear splitting, which is undeformed by the SD radiative background.

%%%%%%%%%%%%%%%%%%%%%
%%%%%%%%%%%%%%%%%%%%%

\section{Gravity: leading holomorphic splitting}\label{sec:holOPEgravity}

In this section, following a road-map similar to Section~\ref{sec:holOPEgauge}, we compute the gravitational leading holomorphic collinear splitting functions from two perspectives. First, by directly evaluating the formula for tree-level MHV graviton scattering on SD radiative spacetimes \eqref{MHVgrav}, and, second, by using graviton vertex operators in the twistor sigma model. There are substantial further complications in gravity compared to Yang-Mills in both approaches. Nevertheless, in both cases we find that the holomorphic splitting functions, and thus the holomorphic celestial OPE coefficients, are identical to those in Minkowski space.

%%%%%%%%%%%%%%%%%%%%%
\subsection{From the MHV amplitude}

Unlike in Yang-Mills theory, the holomorphic collinear limit of MHV amplitudes on a SD radiative spacetime depends significantly on the helicities of the collinear gravitons. We treat the more straightforward case where both gravitons are positive helicity first, dealing with the more complicated mixed-helicity case afterwards.

%%%%%%%%

\subsubsection{Positive helicity with positive helicity}

The holomorphic collinear limit for gravity is parameterized in exactly the same way as in gauge theory, using \eqref{collim} -- \eqref{collms}. In the MHV amplitude \eqref{MHVgrav}, our goal is to isolate contributions which are singular in the holomorphic collinear limit; that is, proportional to $\la i\,j\ra^{-1}$, where $i,j$ are any two positive helicity gravitons. Now, the block decomposition \eqref{HHblockMHV} -- \eqref{TblockMHV} shows that the only entries of $\cH$ which have the holomorphic collinear singularity are $\HH_{ij}$, $\HH_{ji}$, $\HH_{ii}$ and $\HH_{jj}$. 

Using the definition \eqref{reddet}, we are free to choose 
\be\label{grpp1}
\mathrm{det}^{\prime}(\cH)=\frac{|\cH^{j}_{j}|}{\la1\,2\ra^2\,\la1\,j\ra^2\,\la2\,j\ra^2}\,, 
\ee
and then expand the resulting minor along the $i^{\mathrm{th}}$ row, obtaining:
\be\label{grpp2}
\begin{split}
|\cH^{j}_{j}|&=\sum_{k\neq 1,2,j}(-1)^{i+k}\,\HH_{ik}\,|\cH^{ji}_{jk}|+\sum_{\m=1}^{t}(-1)^{i+\m}\,\mathfrak{h}_{i\m}\,|\cH^{ji}_{j\m}| \\
&=\frac{[\![i\,j]\!]}{\la i\,j\ra}\,|\cH^{ij}_{ij}|+O(\varepsilon^0)\,,
\end{split}
\ee
in the holomorphic collinear limit. Now, since
\be\label{framescalegr}
H^{\dot\alpha}{}_{\dot\beta}\!\left(x,\,\frac{\la\xi\,i\ra}{\la\xi\,P\ra}\,\kappa_P\right)=H^{\dot\alpha}{}_{\dot\beta}(x,\kappa_P)\,,
\ee
and using \eqref{dotframe2}, this is further simplified to
\be\label{grpp3}
|\cH^{j}_{j}|=\frac{[i\,j]}{\la i\,j\ra}\,|\cH^{ij}_{ij}|+O(\varepsilon^0)\,,
\ee
with un-dressed square brackets in the pre-factor. 

The only remaining dependence on the two collinear momenta in $|\cH^{ij}_{ij}|$ is through the diagonal entries. Now, we have the identity
\begin{multline}\label{grpp4}
\frac{[\![k\,i]\!]\,\la1\,i\ra\,\la2\,i\ra}{\la k\,i\ra\,\la1\,k\ra\,\la2\,k\ra}+\frac{[\![k\,j]\!]\,\la1\,j\ra\,\la2\,j\ra}{\la k\,j\ra\,\la1\,k\ra\,\la2\,k\ra} \\
=\frac{[\![k\,P]\!]\,\la1\,P\ra\,\la2\,P\ra}{\la k\,i\ra\,\la1\,k\ra\,\la2\,k\ra}\,\frac{(\la\xi\,i\ra\,[\tilde{\xi}\,i]+\la\xi\,j\ra\,[\tilde{\xi}\,j])}{\la\xi\,P\ra\,[\tilde{\xi}\,P]}+O(\varepsilon)=\frac{[\![k\,P]\!]\,\la1\,P\ra\,\la2\,P\ra}{\la k\,i\ra\,\la1\,k\ra\,\la2\,k\ra}+O(\varepsilon)\,,
\end{multline}
for any $k\neq i,j$ as a consequence of \eqref{collim} and \eqref{collms}. This implies that
\begin{multline}\label{grpp5}
 \HH_{kk}=\sum_{l\neq i,j,k}\frac{[\![k\,l]\!]}{\la k\,l\ra}\,\frac{\la1\,l\ra\,\la 2\,l\ra}{\la1\,k\ra\,\la2\,k\ra}+\frac{[\![k\,P]\!]}{\la k\,P\ra}\,\frac{\la1\,P\ra\,\la 2\,P\ra}{\la1\,k\ra\,\la2\,k\ra} \\
 -\im\,\sum_{\m=1}^{t}\,\frac{\varepsilon_\m\,[\![k\,\bar{\lambda}_\m]\!]}{\la\,k\,\lambda_\m\ra}\,\frac{\la1\,\lambda_\m\ra\,\la2\,\lambda_\m\ra}{\la1\,k\ra\,\la2\,k\ra}+O(\varepsilon)\,,
\end{multline}
and 
\begin{multline}\label{grpp6}
\T_{\m\m}=-\im\,\varepsilon_\m\sum_{k\neq i,j}\frac{[\![\bar{\lambda}_\m\,k]\!]}{\la\lambda_\m\,k\ra}\,\frac{\la1\,k\ra\,\la2\,k\ra}{\la1\,\lambda_\m\ra\,\la2\,\lambda_\m\ra}-\im\,\varepsilon_\m\,\frac{[\![\bar{\lambda}_\m\,P]\!]}{\la\lambda_\m\,P\ra}\,\frac{\la1\,P\ra\,\la2\,P\ra}{\la1\,\lambda_\m\ra\,\la2\,\lambda_\m\ra}  \\
-\varepsilon_\m\,\sum_{\mathrm{n}\neq\m}\,\frac{\varepsilon_{\mathrm{n}}\,[\![\bar{\lambda}_\m\,\bar{\lambda}_{\mathrm{n}}]\!]}{\la\lambda_\m\,\lambda_\mathrm{n}\ra}\,\frac{\la1\,\lambda_\mathrm{n}\ra\,\la2\,\lambda_{\mathrm{n}}\ra}{\la1\,\lambda_\m\ra\,\la2\,\lambda_\m\ra}+O(\varepsilon)\,,
\end{multline}
which in turn means that we can write
\be\label{grpp7}
|\cH^{j}_{j}|=\frac{[i\,j]}{\la i\,j\ra}\,|\hat{\cH}^{P}_{P}|+O(\varepsilon^0)\,,
\ee
where $\hat{\cH}$ is the matrix for the $(n-1)$-point MHV amplitude where the positive helicity external gravitons $i,j$ have been replaced by a single positive helicity external graviton $P$. Taking into account the remaining factors in \eqref{grpp1}, this gives
\be\label{grpp8}
\mathrm{det}^{\prime}(\cH)=\frac{[i\,j]}{\la i\,j\ra}\,\frac{\la\xi\,P\ra^4}{\la\xi\,i\ra^2\,\la\xi\,j\ra^2}\,\mathrm{det}^{\prime}(\hat{\cH})+O(\varepsilon)\,,
\ee
in the holomorphic collinear limit.

\medskip

There are two remaining ingredients in the collinear limit of \eqref{MHVgrav} which need to be determined. The first is the behaviour of the exponential factor
\be\label{grpp9}
\exp\!\left(\im\sum_{k=1}^{n}\msf{F}^{\dot\alpha}(x,\kappa_k)\,\tilde{\kappa}_{k\,\dot\alpha}\right)\,.
\ee
This is easily obtained using the collinear parametrization \eqref{collim} -- \eqref{collms} and the fact that $\msf{F}^{\dot\alpha}(x,\kappa)$ is homogeneous of degree $+1$ in its second argument, giving:
\be\label{grpp10}
\exp\!\left(\im\sum_{k\neq i,j}\msf{F}^{\dot\alpha}(x,\kappa_k)\,\tilde{\kappa}_{k\,\dot\alpha}+\im\,\msf{F}^{\dot\alpha}(x,\kappa_P)\,\tilde{\kappa}_{P\,\dot\alpha}\right)\,,
\ee
as desired.

The second ingredient is the counting of the background news function contributions. The $n$-point MHV amplitude has at most $n-4$ insertions of the news function, while the $(n-1)$-point amplitude has only $n-5$. So to obtain a meaningful collinear limit, the `maximal news' (i.e., $t=n-4$) contribution to the $n$-point amplitude must have no collinear singularity. 

To see this, observe that for the $t=n-4$ term in \eqref{MHVgrav} must have $p_\m=3$ for each $\m=1\,\dots,n-4$; this can be shown by a simple inductive argument. The differentiation with respect to the formal $\varepsilon_\m$ parameters has the role of extracting from $\mathrm{det}^{\prime}(\cH)$ only those terms which are cubic in $\varepsilon_\m$ for each $\m$. This process completely saturates the reduced determinant, as a consequence of the maximality of $t=n-4$. That is, if there were some minor of the block $\HH$ remaining after extracting the cubic terms in each $\varepsilon_\m$, then it would be possible to add an additional news function insertion and $t=n-4$ would not be maximal, which is a contradiction.

In other words, integrand of the $t=n-4$ term in $\cM_{n,1}$ can only be composed of products of entries of $\cH$ whose derivative with respect to the formal parameters $\varepsilon_\m$ are non-zero. The only such entries are from the blocks $\mathfrak{h}$ and $\mathbb{T}$, and from the diagonal entries of $\HH$. The two blocks $\mathfrak{h}$ and $\mathbb{T}$ have no holomorphic collinear singularities, while the entries $\HH_{ii},\,\HH_{jj}$ do have collinear singularities. However,
\be\label{grpp11}
\frac{\partial\HH_{ii}}{\partial\varepsilon_\m}=-\im\,\frac{[\![i\,\bar{\lambda}_\m]\!]}{\la\,i\,\lambda_\m\ra}\,\frac{\la1\,\lambda_\m\ra\,\la2\,\lambda_\m\ra}{\la1\,i\ra\,\la2\,i\ra}\,,
\ee
which is free from holomorphic collinear singularities, and similarly for $\HH_{jj}$. Thus, the maximal term $t=n-4$ drops out of the holomorphic collinear limit and the counting of news function insertions behaves appropriately.

\medskip

Pulling all of these pieces together, we find that
\begin{multline}\label{grpp12}
\lim_{\varepsilon\to0}\cM_{n,1}(1^-,2^-,\ldots,i^+,\ldots,j^+,\ldots,n^+)\\
=\mathrm{Split}(i^+_2,j^+_2\to P^+_2)\,\cM_{n-1,1}(1^-,2^-,\ldots,P^+,\ldots,n^+)\,,
\end{multline}
with the holomorphic collinear splitting function
\be\label{grppsplit}
\mathrm{Split}(i^+_2,j^+_2\to P^+_2)=\frac{[i\,j]}{\la i\,j\ra}\,\frac{\la\xi\,P\ra^4}{\la\xi\,i\ra^2\,\la\xi\,j\ra^2}\,,
\ee
matching the known result for tree-level collinear splitting of positive helicity gravitons in Minkowski spacetime~\cite{Bern:1998sv,White:2011yy,Akhoury:2011kq}.

Likewise, Mellin transforming this result leads to the leading, singular holomorphic OPE coefficient for positive helicity outgoing gravitons:
\be\label{grppOPE}
C(\Delta_i^+,\,\Delta_j^+; (\Delta_i+\Delta_j)^+|\bar{z}_{ij})=\sum_{m=0}^{\infty}B(\Delta_i+m-1,\Delta_j-1)\,\frac{\bar{z}_{ij}^{m+1}}{m!}\,\dbar_j^m\,,
\ee
again matching the result for Minkowski space~\cite{Pate:2019lpp,Guevara:2021abz}.

%%%%%%%%%%%%%%%%%%%%

\subsubsection{Positive helicity with negative helicity}

While we have obtained the holomorphic collinear limit for two positive helicity gravitons directly from \eqref{MHVgrav}, the mixed-helicity case (where one collinear graviton is positive helicity and the other is negative helicity) is much more subtle. The basic reason for this is the underlying chirality of the background SD radiative spacetime. Even in Minkowski spacetime, where \eqref{MHVgrav} reduces to the Hodges formula~\cite{Hodges:2012ym}, the resulting expression is not manifestly parity-invariant, so it is instructive to first recall how the mixed helicity holomorphic collinear limit is obtained there.

Recall that the Hodges formula for MHV graviton scattering in Minkowski spacetime is
\be\label{Hodges1}
\cM_{n,1}^{\text{flat}}=(2\pi)^4\,\delta\!\left(\sum_{i=1}^{n}k_i\right)\la1\,2\ra^8\,\mathrm{det}^{\prime}(\HH)\,,
\ee
with the flat spacetime entries of $\HH$ given by
\be\label{flatHH}
\HH_{ij}=-\frac{[i\,j]}{\la i\,j\ra}\,, \quad i\neq j\,, \qquad \HH_{ii}=\sum_{j\neq i}\frac{[i\,j]}{\la i\,j\ra}\frac{\la a\,j\ra\,\la b\,j\ra}{\la a\,i\ra\,\la b\,i\ra}\,,
\ee
with the constant reference spinors $a^{\alpha},\,b^{\alpha}$ arbitrarily chosen as a consequence of momentum conservation. In particular, if we were to choose a different reference spinor, $a^{\prime\alpha}$ then a straightforward application of the Schouten identity shows that
\be\label{flatHH2}
\begin{split}
\Delta\HH_{ii}&=\sum_{j\neq i}\frac{[i\,j]}{\la i\,j\ra}\,\frac{\la b\,j\ra}{\la a\,i\ra\,\la a'\,i\ra\,\la b\,i\ra}\left(\la a\,j\ra\,\la a'\,i\ra-\la a'\,j\ra\,\la a\,i\ra\right) \\
 &=-\frac{\la a\,a'\ra}{\la a\,i\ra\,\la a'\,i\ra\,\la b\,i\ra}\sum_{j=1}^{n}[i\,j]\,\la b\,j\ra=0\,,
\end{split}
\ee
by momentum conservation.

Now, let $i$ be any one of the positive helicity gravitons, and consider the holomorphic collinear limit between $1$ and $i$. Using the definition of $\mathrm{det}^{\prime}(\HH)$ and the usual parametrization of the collinear limit, it follows that 
\be\label{flatcoll1}
\begin{split}
\mathrm{det}^{\prime}(\HH)&=\frac{|\HH^{j}_{j}|}{\la1\,2\ra^{2}\,\la1\,j\ra^2\,\la2\,j\ra^2} =\frac{1}{\la1\,2\ra^{2}\,\la1\,j\ra^2\,\la2\,j\ra^2}\sum_{k\neq1,2,j}(-1)^{i+k}\,\HH_{ik}\,|\HH^{ij}_{kj}| \\
&=\frac{[1\,i]}{\la1\,i\ra}\,\frac{\la\xi\,P\ra^{4}}{\la\xi\,1\ra^2\,\la\xi\,i\ra^2}\,\frac{|\HH^{ij}_{ij}|}{\la P\,2\ra^2\,\la P\,j\ra^2\,\la2\,j\ra^2}+O(\varepsilon^0) \\
&=\frac{[1\,i]}{\la1\,i\ra}\,\frac{\la\xi\,P\ra^{4}}{\la\xi\,1\ra^2\,\la\xi\,i\ra^2}\,\mathrm{det}^{\prime}(\hat{\HH})+O(\varepsilon^0)\,,
\end{split}
\ee
where $\hat{\HH}$ is the $(n-3)\times(n-3)$ Hodges matrix with the negative helicity graviton $1$ and positive helicity graviton $i$ replaced by the negative helicity collinear graviton $P$. Combined with the collinear limit of the overall factor of $\la1\,2\ra^8$, this gives the well-known mixed-helicity holomorphic collinear splitting~\cite{Bern:1998sv,White:2011yy,Akhoury:2011kq}
\be\label{flatcoll2}
\lim_{\varepsilon\to0}\cM_{n,1}^{\text{flat}}(1^-,\ldots,i^+,\ldots)=\frac{[1\,i]}{\la1\,i\ra}\,\frac{\la\xi\,1\ra^6}{\la\xi\,P\ra^4\,\la\xi\,i\ra^2}\,\cM_{n-1,1}^{\text{flat}}(P^-,\ldots)\,.
\ee
This relied crucially on the ability to select arbitrary reference spinors in the diagonal entries of $\HH$, which meant that the holomorphic collinear singularity $\la1\,i\ra^{-1}$ appears as a single term in the diagonal entry $\HH_{ii}$. If we set $a=1$, $b=2$ this term is removed and the correct collinear splitting function would only emerge after careful application of momentum conservation in the collinear limit.

\medskip

For MHV graviton scattering in a SD radiative spacetimes, the reference spinors in formula \eqref{MHVgrav} are fixed to be the negative helicity gravitons. This is an artefact of the formula's derivation~\cite{Adamo:2022mev}, which relies on expressing the MHV generating functional on spacetime using an ASD spin-frame adapted to the two negative helicity gravitons. A na\"ive extension of \eqref{MHVgrav} to include arbitrary reference spinors in the diagonal entries fails: with no momentum conservation, the resulting formula depends on the choice of these spinors, which is clearly un-physical (see Appendix~\ref{appA}).

Consequently, other strategies are required to tackle the mixed-helicity case at the level of the amplitude itself. To do this, we adopt two complementary strategies: first, we compute the holomorphic collinear limit directly from the 4-point amplitude computed using Feynman rules defined by the background field formalism. For this calculation to be tractable, we work with a particularly simple subset of SD radiative metrics, namely self-dual plane waves. Second, we show that a conjectural generalization of \eqref{MHVgrav} with arbitrary reference spinors (defined for the full tree-level S-matrix in appendix~\ref{appA}) also produces the holomorphic collinear splitting function on general SD radiative spacetimes and at arbitrary multiplicity.

\paragraph{Feynman diagrams:} We begin by computing the holomorphic collinear limit directly from background-coupled Feynman rules on a particular SD radiative background where such computations are actually tractable: a \emph{self-dual plane wave} (SDPW). These spacetimes have a covariantly constant null Killing vector $n^{\alpha\dot\alpha}=\iota^{\alpha}\,\tilde{\iota}^{\dot\alpha}$, and are defined by the radiative data~\cite{Ward:1978soq,Curtis:1978pw,Porter:1982uj}
\be\label{SDPWrad}
\tilde{\sigma}^0(u,\lambda,\bar{\lambda})\,\D\bar{\lambda}=\frac{\la o\,\lambda\ra^2}{[\tilde{o}\,\bar{\lambda}]}\,\bar{\delta}(\la\iota\,\lambda\ra)\,\cF\!\left(\frac{u}{\la\lambda\,o\ra\,[\bar{\lambda}\,\tilde{o}]}\right)\,,
\ee
where $\{o_{\alpha},\,\tilde{o}_{\dot\alpha}\}$ are constant spinors normalized to $\la\iota\,o\ra=1=[\tilde{\iota}\,\tilde{o}]$. In lightfront coordinates where $\iota^{\alpha}\tilde{\iota}^{\dot\alpha}\partial_{\alpha\dot\alpha}=\partial_{+}$, the free function
\be\label{SDPWmet}
\cF(x^-)=\int^{x^-}\!\d s\,f(s)\,, \qquad \tilde{\Psi}_{\dot\alpha\dot\beta\dot\gamma\dot\delta}=-\ddot{f}(x^-)\,\tilde{\iota}_{\dot\alpha}\tilde{\iota}_{\dot\beta}\tilde{\iota}_{\dot\gamma}\tilde{\iota}_{\dot\delta}\,,
\ee
for $f(x^-)$ the profile function entering the SDPW metric and $\tilde{\Psi}_{\dot\alpha\dot\beta\dot\gamma\dot\delta}$ the self-dual Weyl curvature spinor.

The 4-point graviton amplitude in these spacetimes was computed directly from the background field Lagrangian in~\cite{Adamo:2022mev}, with contributions from four distinct Feynman diagrams:
\begin{equation}
    \mathcal{M}_{4,1} = \mathcal{M}_{\rm{s}} + \mathcal{M}_{\rm{t}}+ \mathcal{M}_{\rm{u}} + \mathcal{M}_{\rm{cont}} \, ,
\end{equation}
corresponding to the three exchange channels (labeled by their Mandelstam variables) and the 4-graviton contact interaction. Working in a hybrid gauge adapted to the Killing vector $n^{\alpha\dot\alpha}$ of the SDPW metric and the momentum of one of the negative helicity gravitons, the (background-dressed) graviton polarization tensors are:
\be\label{neggravpol}
\mathcal{E}_{i}^{\alpha\dot\alpha\beta\dot\beta}=\frac{\kappa_i^{\alpha}\,\kappa_i^{\beta}\,\tilde{\iota}^{\dot\alpha}\tilde{\iota}^{\dot\beta}}{[\tilde{\iota}\,i]^2}\,, \qquad i=1,2\,,
\ee
\be\label{posgravpol}
\mathcal{E}_{j}^{\alpha\dot\alpha\beta\dot\beta}=\frac{\kappa_2^{\alpha}\,\kappa_2^\beta}{\la2\,j\ra^2}\left(\tilde{K}_j^{\dot\alpha}\,\tilde{K}_j^{\dot\beta}-\im\,\dot{f}\,\frac{[\tilde{\iota}\,j]}{\la\iota\,j\ra}\,\tilde{\iota}^{\dot\alpha}\,\tilde{\iota}^{\dot\beta}\right)\,, \qquad j=3,4\,.
\ee
Here, all four gravitons are in lightfront-de Donder gauge, with the lightfront vector for the negative helicity gravitons being $n^{\alpha\dot\alpha}$ while for the positive helicity gravitons it is $k_{2}^{\alpha\dot\alpha}$. In this gauge, the contact diagram $\cM_{\mathrm{cont}}=0$, and only the exchanges contribute.

Let us consider the holomorphic collinear limit $\la1\,3\ra\to0$. It is straightforward to show that the only contribution to the 4-point amplitude which is potentially singular in this limit is from the t-channel. The full expression for the t-channel diagram in our chosen gauge\footnote{Note that this expression is not symmetric under interchange of the negative helicity gravitons $1\leftrightarrow2$ as a result of the gauge choice, which singles out $\kappa_{2\,\alpha}$ to define lightfront gauge for the positive helicity gravitons. This is to be expected, as $\cM_{\mathrm{t}}$ is just a single Feynman diagram, and need not be gauge invariant or permutation symmetric on its own; the full 4-point amplitude, including s- and u-channels, \emph{does} have these properties, as required~\cite{Adamo:2022mev}.} is easily obtained from~\cite{Adamo:2022mev}: 
\begin{equation}
\begin{split}
     \mathcal{M}_{\rm{t}} &= \delta^3_{+, \perp}\left( \sum_{i=1}^4 k_i \right) \frac{\langle 12 \rangle^3}{\langle 23 \rangle} \frac{[\tilde{\iota}\, 4]^2}{(k_2+k_4)_+^2} \\
     &\times \int \d^2 \mu[\mathrm{t}]\, \Bigg[  \frac{\langle \iota \, 2 \rangle^2}{[\tilde{\iota} \, 2]^2} \frac{[\tilde{\iota} \, 3]^3}{[\tilde{\iota}\, 1]}\Bigg( \llbracket 24\rrbracket^2 - \mathrm{i}\,  (k_2+k_4)_+ \frac{[\tilde{\iota}\, 2]}{\langle \iota \, 2 \rangle} \frac{[\tilde{\iota}\, 4]}{\langle \iota \, 4 \rangle} \, \dot{f} \Bigg) (x^-) \\
     &~+~ \frac{\langle \iota \, 1 \rangle}{[\tilde{\iota} \, 1] } \frac{[\tilde{\iota} \, 4]^2 \{\langle \iota \, 2 \rangle[\tilde{\iota}\, 3]\langle 31\rangle + \langle \iota \, 1 \rangle [\tilde{\iota}\, 1]\langle 21\rangle \}}{\langle 32 \rangle\,  [\tilde{\iota}\, 2]^2} \Bigg( \llbracket 13\rrbracket^2 - \mathrm{i}\,  (k_1+k_3)_+ \frac{[\tilde{\iota}\, 1]}{\langle \iota \, 1 \rangle} \frac{[\tilde{\iota}\, 3]}{\langle \iota \, 3 \rangle} \, \dot{f} \Bigg) (y^-) \\
     &+ \frac{\langle \iota \, 2 \rangle [\tilde{\iota} \, 4] [\tilde{\iota}\, 3]}{\langle 32\rangle [\tilde{\iota} \, 2]^2 [\tilde{\iota} \, 1]} ( \langle \iota \, 2\rangle [\tilde{\iota}\, 3]\langle 31\rangle + \langle \iota \, 1\rangle [\tilde{\iota} \, 1]\langle 21\rangle + \langle \iota\, 1\rangle [\tilde{\iota} \, 3]\langle 32\rangle )\, \llbracket 24 \rrbracket(x^-) \llbracket 13\rrbracket(y^-) \Bigg]\\
     &+ (x^- \leftrightarrow y^-) \, , \label{eq:tch}
\end{split}
\end{equation}
where the measure
\begin{multline}\label{tmeasure}
\d^2 \mu[\mathrm{t}]:=\Theta(x^--y^-)\,\frac{\d x^-\,\d y^-}{(k_1+k_3)_+}\,\exp\!\left[\im\int^{x^-}\!\!\d s\,\Big(\la o\,1\ra\,[\![\tilde{o}\,1]\!](s)+\la o\,3\ra\,[\![\tilde{o}\,3]\!](s)\Big)\right. \\
+\im\int^{y^-}\!\!\d s\,\Big(\la o\,2\ra\,[\![\tilde{o}\,2]\!](s) +\la o\,4\ra\,[\![\tilde{o}\,4]\!](s)\Big) \\
\left.-\im\,\frac{\la o|k_1+k_3|\tilde{\iota}]}{(k_1+k_3)_+}\int_{x^-}^{y^-}\!\d s\,\Big(\la\iota|k_1+k_3|\tilde{o}]+\la o|k_1+k_3|\tilde{\iota}]\,f(s)\Big)\right]\,,
\end{multline}
arises from the structure of the graviton propagator in the SDPW background. After an integration-by-parts which removes one of the lightfront time integrals~\cite{Adamo:2022mev}, the only term in \eqref{eq:tch} with a holomorphic collinear singularity can be put into the form
\begin{multline}\label{tchsing1}
\delta^3_{+, \perp}\left(\sum_{i=1}^4 k_i \right)\frac{\langle 12\rangle^4 [\tilde{\iota}\, 4]^2}{\langle 32 \rangle\, (k_1+k_3)^2_+} \frac{\langle \iota \, 1\rangle^2\, [\tilde{\iota} \, 4]^2}{\langle 3\,2\rangle\, [\tilde{\iota}\, 2]^2}\int\d x^-\,\frac{\llbracket 1\,3 \rrbracket(x^-)}{\langle 1\,3\rangle} \\
\times \exp\!\left[\im\sum_{j=1}^{4}\int^{x^-}\!\!\d s\,\la o\,j\ra\,[\![\tilde{o}\,j]\!](s)\right]\,,
\end{multline}
where the available momentum conserving delta functions have been used.

Thus, we find that the holomorphic collinear limit, with the usual parametrization \eqref{collim}, \eqref{collms}, is given by
\begin{multline}\label{tchsing2}
\lim_{\varepsilon\to0}\cM_{4,1}=\delta^3_{+, \perp}\left(P+k_2+k_4\right)\frac{[1\,3]}{\la1\,3\ra}\,\frac{\la\xi\,1\ra^6}{\la\xi\,P\ra^4\,\la\xi\,3\ra^2}\,\frac{\la P\,2\ra^2\,[\tilde{\iota}\,4]^4}{[\tilde{\iota}\,P]^2\,[\tilde{\iota}\,2]^2} \\
\times \int\d x^-\,\exp\!\left[\im\int^{x^-}\!\!\d s\left(\sum_{j=2,4}\la o\,j\ra\,[\![\tilde{o}\,j]\!](s)+\la o\,P\ra\,[\![\tilde{o}\,P]\!](s)\right)\right]\,.
\end{multline}
This can be further simplified using the available momentum conservation, which implies that
\be\label{tchsing3}
\la2\,4\ra\,[4\,\tilde{\iota}]=-\la2\,P\ra\,[P\,\tilde{\iota}]\,, \qquad \la4\,2\ra\,[2\,\tilde{\iota}]=-\la4\,P\ra\,[P\,\tilde{\iota}]\,, \qquad \la P\, 4\ra \,[4 \, \tilde{\iota}]= -\la P\, 2\ra [2\, \tilde{\iota}] \, .
\ee
This gives
\begin{multline}\label{tchsing4}
\lim_{\varepsilon\to0}\cM_{4,1}=\delta^3_{+, \perp}\left(P+k_2+k_4\right)\frac{[1\,3]}{\la1\,3\ra}\,\frac{\la\xi\,1\ra^6}{\la\xi\,P\ra^4\,\la\xi\,3\ra^2}\,\frac{\la P\,2\ra^6}{\la P\,4\ra^2\,\la2\,4\ra^2} \\
\times \int\d x^-\,\exp\!\left[\im\int^{x^-}\!\!\d s\left(\sum_{j=2,4}\la o\,j\ra\,[\![\tilde{o}\,j]\!](s)+\la o\,P\ra\,[\![\tilde{o}\,P]\!](s)\right)\right] \\
=\mathrm{Split}(1^-_2,3^+_2\to P^-_2)\,\cM_{3,1}\,,
\end{multline}
where 
\be\label{grpnsplit}
\mathrm{Split}(1^-_2,3^+_2\to P^-_2)=\frac{[1\,3]}{\la1\,3\ra}\,\frac{\la\xi\,1\ra^6}{\la\xi\,P\ra^4\,\la\xi\,3\ra^2}\,,
\ee
is the same mixed-helicity holomorphic collinear splitting function that emerged in \eqref{flatcoll2} on Minkowski space.

\medskip

\paragraph{Generalised amplitude formula:} While this Feynman diagram calculation on a SDPW metric is very explicit, one could worry that the special, highly symmetric choice of SD radiative background and restriction to 4-points -- with the resulting, highly constrained 3-particle kinematics of the collinear limit -- make for a result which is not sufficiently robust. As explained at the beginning of this subsection, to give a more general argument at the level of an all-multiplicity MHV amplitude, a formula with general reference spinors on the SD radiative background is needed.

While we do not know a way to derive such a formula from first principles, one with the desired property of being independent of the choice of reference spinors can be guessed using some basic guidelines. In Appendix~\ref{appA}, we present this generalized formula for arbitrary N$^{k}$MHV degree, and show why it is independent of the choice of reference spinors. Here, we simply present the expression at MHV:
\be\label{genMHV}
\cM_{n,1}=\sum_{\gamma=0}^{2n-7}C^{(\gamma)}_{n,1}\,,
\ee
where
\begin{multline}\label{genMHV1}
C^{(\gamma)}_{n,1}=\frac{(-1)^{\gamma}}{\gamma!}\,\la1\,2\ra^8 \sum_{t=0}^{n-4}\sum_{p_1,\ldots,p_t>2}\int\d^{4}x\,\sqrt{g}\,\left(\prod_{\m=1}^{t}\,\frac{\partial^{p_\m}}{\partial\varepsilon_\m^{p_\m}}\right) \\
\sum_{\beta=0}^{\gamma} \sum_{\substack{i_1,\ldots,i_b \\ \m_1,\ldots,m_{\gamma-\beta}}}f_{i_1}^{\alpha_1}\cdots f_{i_\beta}^{\alpha_\beta}\,f_{\m_1}^{\alpha_{\beta+1}}\cdots f_{\m_{\gamma-\beta}}^{\alpha_\gamma}\\
\frac{\partial}{\partial x^{\alpha_1\dot\alpha_1}}\cdots\frac{\partial}{\partial x^{\alpha_\gamma\dot\alpha_\gamma}}\left(\tilde{K}_{i_1}^{\dot\alpha_1}\cdots\tilde{K}_{i_\beta}^{\dot\alpha_\beta}\,\bar{\Lambda}^{\dot\alpha_{\beta+1}}_{\m_1}\cdots\bar{\Lambda}^{\dot\alpha_\gamma}_{\m_{\gamma-\beta}}\,\mathrm{det}'(\cG^{i_1\cdots i_{\beta}\m_1\cdots\m_{\gamma-\beta}}_{i_1\cdots i_{\beta}\m_1\cdots\m_{\gamma-\beta}})\right)\Big|_{\varepsilon_{\m}=0}\\
 \times \exp\!\left(\im\sum_{i=1}^{n}\msf{F}^{\dot\alpha}(x,\kappa_i)\,\tilde{\kappa}_{i\,\dot\alpha}\right)\,\prod_{\m=1}^{t}\D\lambda_\m\wedge\D\bar{\lambda}_\m\,N^{(p_\m-2)}(\lambda_\m)\,,
\end{multline}
where
\be\label{fMHV}
f^{\alpha}_{i}:=\frac{\im}{2\,\la a\,i\ra\,\la b\,i\ra}\left(\frac{\la a\,k\ra\,b^{\alpha}}{\la k\,i\ra}+\frac{\la a\,l\ra\,b^{\alpha}}{\la l\,i\ra}+\frac{\la b\,k\ra\,a^{\alpha}}{\la k\,i\ra}+\frac{\la b\,l\ra\,a^{\alpha}}{\la l\,i\ra}\right)\,
\ee
Here, $a^{\alpha},b^{\alpha}$ are arbitrary reference spinors, and $k,l$ are \emph{any} two distinct external gravitons. The $(n+t-2)\times(n+t-2)$ matrix $\cG$ has the same block decomposition, differing from $\cH$ only on the diagonal:
\be\label{modHH*}
\HH_{ii}\to\sum_{j\neq i}\,\frac{[\![i\,j]\!]}{\la i\,j\ra}\,\frac{\la a\,j\ra\,\la b\,j\ra}{\la a\,i\ra\,\la b\,i\ra}-\im\sum_{\m=1}^{t}\,\frac{\varepsilon_\m\,[\![i\,\bar{\lambda}_\m]\!]}{\la i\,\lambda_\m\ra}\,\frac{\la a\,\lambda_\m\ra\,\la b\,\lambda_\m\ra}{\la a\,i\ra\,\la b\,i\ra}\,,
\ee
\be\label{modTT*}
\T_{\m\m}\to-\im\,\varepsilon_\m\sum_{i=1}^{n}\,\frac{[\![\bar{\lambda}_\m\,i]\!]}{\la\lambda_\m\,i\ra}\,\frac{\la a\,i\ra\,\la b\,i\ra}{\la a\,\lambda_\m\ra\,\la b\,\lambda_\m\ra}-\varepsilon_\m\sum_{\mathrm{l}\neq\m}\,\frac{\varepsilon_{\mathrm{l}}\,[\![\bar{\lambda}_\m\,\bar{\lambda}_{\mathrm{l}}]\!]}{\la\lambda_\m\,\lambda_\mathrm{l}\ra}\,\frac{\la a\,\lambda_{\mathrm{l}}\ra\,\la b\,\lambda_{\mathrm{l}}\ra}{\la a\,\lambda_\m\ra\,\la b\,\lambda_\m\ra}\,.
\ee
Crucially, this means that for generic $a^{\alpha}, b^{\alpha}$, the holomorphic collinear pole between the negative helicity graviton 1 and any positive helicity graviton $i$ appears in the diagonal entries of the $\HH$ block in exactly the same way as in Minkowski space.

Indeed, the behaviour of the rest of the matrix $\cG$ in the collinear limit is the same as in the case where both gravitons have positive helicity, so it follows immediately that
\be\label{grpn1}
\lim_{\varepsilon\to0}C^{(0)}_{n,1}(1^-,2^-\ldots,i^+,\ldots)=\mathrm{Split}(1^-_2,i^+_2\to P^-_2)\,C^{(0)}_{n-1,1}(P^-,2^-,\ldots)\,.
\ee
All other terms in \eqref{genMHV} will likewise have the correct collinear splitting function, emerging from the further reduced minors of $\cG$, and the counting of tail terms will work precisely as it did in the same helicity case. The only potential issue is the presence of new collinear divergences in $f_{i}^{\alpha}$ appearing in the $\gamma>0$ terms. However, by taking $k=2$ and $l=j$, for any $j\neq i$, it is immediately clear that $f_i^{\alpha}$ is regular in the holomorphic collinear limit, so no such additional divergences arise.

In other words, the formula \eqref{genMHV} obeys
\be\label{grpn2}
\lim_{\varepsilon\to0}\cM_{n,1}(1^-,2^-\ldots,i^+,\ldots)=\mathrm{Split}(1^-_2,i^+_2\to P^-_2)\,\cM_{n-1,1}(P^-,2^-,\ldots)\,,
\ee
as desired.

%%%%%%%%%%%%%%%%%%%%%%%%%%
%%%%%%%%%%%%%%%%%%%%%%%%%%

\subsection{From the twistor sigma model}

We now compute the collinear splitting functions using graviton vertex operators in the twistor sigma model \eqref{tsigm}. As we saw in the gauge theory case of Section~\ref{subsection_YM_TSM_calculation}, this calculation can be performed for generic SD radiative backgrounds and at arbitrary N$^{d-1}$MHV degree. In the twistor sigma model, Wick contractions between vertex operators are governed by \eqref{mprop}:
\be\label{mprop*}
m^{\dot\alpha}(\sigma)\,m^{\dot\beta}(\sigma')=\frac{H^{\dot\alpha}{}_{\dot\gamma}(\sigma)\,H^{\dot\beta\dot\gamma}(\sigma')}{(\sigma\,\sigma')}\,\mathfrak{s}_{\tilde{\texttt{h}}}(\sigma)\,\mathfrak{s}_{\tilde{\texttt{h}}}(\sigma')\,,
\ee
where $\mathfrak{s}_{\tilde{\texttt{h}}}$ is defined by \eqref{sdef} in terms of $\tilde{\mathtt{h}}$, the set of $d+1$ points on $\P^1$ corresponding to the negative helicity gravitons in a N$^{d-1}$MHV scattering process.

The contraction between two graviton vertex operators should, in principle, also be dressed by the inclusion of arbitrarily many background vertex operators \eqref{backVO}. These vertex operators represent background interactions in the twistor sigma model, and as such can contribute to any contraction or correlator\footnote{For instance, the calculation of beta functionals in background-coupled string theory can be performed in this way~\cite{Kato:1982im,Friedan:1985ii,Banks:1986fu}.}. However, the contribution of these background vertex operators to the contraction between two graviton vertices clearly involves multiple Wick contractions: at least one must be between the graviton vertices themselves (by definition) so interactions with the background vertices must entail at least one additional contraction. Thus, the classical contraction between graviton vertex operators, which must involve only a single Wick contraction, does not involve contributions from the background vertex operators.

\medskip

We begin with the case of two positive helicity gravitons, which are described by the vertex operators \eqref{gravVO} with momentum eigenstate representatives \eqref{gravmomeig}. In particular, we have:
\begin{equation}\label{Graviton_vo}
    \cU_{+\,i} = \int\frac{\D\sigma_i}{\mathfrak{s}^2_{\tilde{\texttt{h}}}(\sigma_i)}\,\frac{\d t_i}{t_i^3}\,\bar\delta^2(\kappa_i-t_i\,\lambda(\sigma_i))\, \e^{\im\, t_i\,[\bmu(\sigma_i)\,i]}\,,
\end{equation}
and similarly for $\cU_{+\,j}$, where $\bmu^{\dot\alpha}(\sigma)=\msf{F}^{\dot\alpha}(\sigma)+m^{\dot\alpha}(\sigma)$ for $\msf{F}^{\dot\alpha}(\sigma)$ giving the degree $d$ rational curve in twistor space, holomorphic with respect to the complex structure \eqref{SDradmet}, $\bar{\nabla}$. Thus, all Wick contractions between graviton vertex operators via \eqref{mprop*} are controlled by the contractions between the $m^{\dot\alpha}$-dependent exponentials.

Using the standard formula for Wick contractions between exponential functions (cf., \cite{Polchinski:1998rq}), one obtains
\be\label{expcont}
\e^{\im\,t_i\,[m(\sigma_i)\,i]}\,\e^{\im\,t_j\,[m(\sigma_j)\,j]}=\exp\!\left(-\frac{t_i\,t_j\,[\![i\,j]\!]}{(i\,j)}\,\mathfrak{s}_{\tilde{\texttt{h}}}(\sigma_i)\,\mathfrak{s}_{\tilde{\texttt{h}}}(\sigma_j)\right)\,:\,\e^{\im\,t_i\,[m(\sigma_i)\,i]+\im\,t_j\,[m(\sigma_j)\,j]}\,:\,,
\ee
where `$:\cdots:$' denotes normal-ordering. The classical contribution corresponds to taking the linear term in the first exponential on the right-hand-side of this expression, encoding a single Wick contraction. Hence, the classical contraction between positive helicity graviton vertex operators is given by:
\begin{multline}\label{grws1}
\cU_{+\,i}\,\cU_{+\,j}\Big|_{\text{class.}}=-\int\frac{\D\sigma_i\,\D\sigma_j}{\mathfrak{s}_{\tilde{\texttt{h}}}(\sigma_i)\,\mathfrak{s}_{\tilde{\texttt{h}}}(\sigma_j)}\,\frac{[\![i\,j]\!]}{(i\,j)}\,\frac{\d t_i}{t_i^2}\,\frac{\d t_j}{t_j^2} \\
\bar{\delta}^{2}(\kappa_i-t_i\,\lambda(\sigma_i)) \,\bar{\delta}^{2}(\kappa_j-t_j\,\lambda(\sigma_j))\,\exp\!\left[\im\left(t_i\,[\bmu(\sigma_i)\,i]+t_j\,[\bmu(\sigma_j)\,j]\right)\right]\,.
\end{multline}
At this point, the calculation largely follows the lines of that for gluon vertex operators in Section~\ref{subsection_YM_TSM_calculation}. Performing the integral in $t_i$ explicitly leaves
\begin{multline}\label{grws2}
-\int\frac{\D\sigma_i\,\D\sigma_j}{\mathfrak{s}_{\tilde{\texttt{h}}}(\sigma_i)\,\mathfrak{s}_{\tilde{\texttt{h}}}(\sigma_j)}\,\frac{[\![i\,j]\!]}{(i\,j)}\,\frac{\d t_j}{t_j^2}\,\frac{\la\xi\,\lambda(\sigma_i)\ra^2}{\la\xi\,i\ra^2} \\
\bar{\delta}(\la i\,\lambda(\sigma_i)\ra) \,\bar{\delta}^{2}(\kappa_j-t_j\,\lambda(\sigma_j))\,\exp\!\left[\im\left(\frac{[\bmu(\sigma_i)\,i]\,\la\xi\,i\ra}{\la\xi\,\lambda(\sigma_i)\ra}+t_j\,[\bmu(\sigma_j)\,j]\right)\right]\,,
\end{multline}
and the holomorphic delta function $\bar{\delta}(\la i\,\lambda(\sigma_i)\ra)$ can be expressed as
\be\label{grholdel}
\bar{\delta}(\la i\,\lambda(\sigma_i)\ra)=\frac{1}{2\pi\im}\,\bar{\nabla}_i\left(\frac{1}{\la i\,\lambda(\sigma_i)\ra}\right)\,,
\ee
where $\bar{\nabla}_i$ denotes the pullback of the twistor space complex structure \eqref{SDradmet} to the point $\sigma_i$. This holds because $\bar{\nabla}_i$ differs from $\dbar_i$ only be terms which act as $\partial/\partial\mu^{\dot\alpha}(\sigma_i)$, and the argument of the holomorphic delta function is independent of $\mu^{\dot\alpha}(\sigma_i)$.

Since $\bar{\nabla}$ is integrable by definition, we can integrate-by-parts in $\sigma_i$ to act with $\bar{\nabla}_i$ on the rest of the integrand in \eqref{grws2}. Now, using \eqref{dotframe2} and the fact that $m^{\dot\alpha}(\sigma)$ has simple zeros at each $\sigma_k$, $k\in\tilde{\mathtt{h}}$, the only contribution comes from the action of $\bar{\nabla}_i$ on the simple pole in $(i\,j)$. This gives a holomorphic delta function $\bar{\delta}(i\,j)$ against which the $\D\sigma_i$ integral can be performed, setting $\sigma_i\to\sigma_j$ everywhere else in the integrand to give:
\begin{multline}\label{grws3}
\frac{[i\,j]}{\la i\,j\ra}\,\frac{\la\xi\,j\ra^2}{\la\xi\,i\ra^2}\int\frac{\D\sigma_j}{\mathfrak{s}^2_{\tilde{\texttt{h}}}(\sigma_j)}\,\frac{\d t_j}{t_j^3}\,\bar{\delta}^2(\kappa_j-t_j\,\lambda(\sigma_j)) \\
\times\,\exp\!\left[\frac{\im\,t_j}{\la\xi\,j\ra}\left([\bmu(\sigma_i)\,i]\,\la\xi\,i\ra+[\bmu(\sigma_j)\,j]\,\la\xi\,j\ra\right)\right]\,,
\end{multline}
having invoked the support of the remaining holomorphic delta functions. Upon rescaling $t_j\to t_j\frac{\la\xi\,j\ra}{\la\xi\,P\ra}$ and invoking the holomorphic collinear limit \eqref{collim}, \eqref{collms}, we arrive at
\be\label{grws4}
\cU_{+\,i}\,\cU_{+\,j}\Big|_{\text{class.}}=\frac{[i\,j]}{\la i\,j\ra}\,\frac{\la\xi\,P\ra^4}{\la\xi\,i\ra^2\,\la\xi\,j\ra^2}\,\cU_{+\,P}+O(\varepsilon)\,,
\ee
giving the same holomorphic collinear splitting as Minkowski spacetime.

The computation of the mixed-helicity graviton vertex operator contraction proceeds along identical lines, although the negative helicity graviton vertex operator itself must be defined artificially, as it does not arise in the twistor sigma model which describes purely self-dual perturbations on the background. In particular, we must assume that the negative helicity graviton labelled by the vertex operator is \emph{not} one of the entries in the set of $d+1$ points $\tilde{\mathtt{h}}$, and must also invoke the existence of a non-dynamical object $\msf{O}(\sigma)$, valued in $\cO(8)$ on twistor space. These can both be made precise by embedding the negative helicity graviton at the bottom of a $\cN=8$ SUGRA multiplet on twistor space (e.g., \cite{Skinner:2013xp}), although for our purposes it suffices to take
\be\label{ngravvo}
\cU_{-\,i}=\int\frac{\D\sigma_i}{\mathfrak{s}^2_{\tilde{\texttt{h}}}(\sigma_i)}\,\d t_i\,t_i^5\,\msf{O}(\sigma_i)\,\bar\delta^2(\kappa_i-t_i\,\lambda(\sigma_i))\, \e^{\im\, t_i\,[\bmu(\sigma_i)\,i]}\,.
\ee
With this definition, one soon finds
\be\label{grws5}
\cU_{+\,i}\,\cU_{-\,j}\Big|_{\text{class.}}=\frac{[i\,j]}{\la i\,j\ra}\,\frac{\la\xi\,j\ra^6}{\la\xi\,P\ra^4\,\la\xi\,i\ra^2}\,\cU_{-\,P}+O(\varepsilon)\,,
\ee
again giving the same holomorphic collinear splitting function as in Minkowski space.

%%%%%%%%%%%%%%%%%%%%%
%%%%%%%%%%%%%%%%%%%%%

\section{Infinite dimensional chiral algebras} \label{sec5}

In perturbative gauge theory and gravity on trivial backgrounds (i.e., a flat gauge field or Minkowski spacetime), infinitesimal deformations of the self-dual sector form infinite-dimensional chiral algebras~\cite{Guevara:2021abz,Strominger:2021lvk,Himwich:2021dau}. The generators of these algebras are modes for the conformally soft positive helicity gluons and gravitons, and are often also called infinite-dimensional soft algebras. In gauge theory on a trivial background, this chiral algebra is the loop algebra of Lie algebra of polynomial maps from the complex 2-plane to the Lie algebra of the gauge group, $\cL\mathfrak{g}[\C^2]$, sometimes referred to as the $\cS$-algebra. In gravity on Minkowski space, the chiral algebra is the loop algebra of the Lie algebra of holomorphic Hamiltonian vector fields on $\C^2$, $\cL\mathfrak{ham}(\C^2)$, sometimes identified with $\cL w^{\wedge}_{1+\infty}$, the loop algebra of the wedge algebra of $w_{1+\infty}$.

It is known that when SD background fields with sources or `large' data are turned on, these chiral algebras are deformed: this happens in monopole backgrounds~\cite{Garner:2023izn} and in Eguchi-Hansen~\cite{Bittleston:2023bzp} and Burns spaces~\cite{Costello:2022jpg,Costello:2023hmi}, as well as for non-commutative deformations of the spacetime theory~\cite{Bu:2022iak,Monteiro:2022lwm,Monteiro:2022xwq}. However, we will see that in \emph{any} SD radiative background, the infinite-dimensional chiral algebras remain $\cL\mathfrak{g}[\C^2]$ and $\cL\mathfrak{ham}(\C^2)$ in gauge theory and gravity, respectively.

%%%%%%%%%%%%%%%%%%%%

\subsection{Yang-Mills}

An outgoing, positive helicity gluon coupled to a SD radiative background gauge field is represented in the conformal primary basis by a vertex operator~\cite{Adamo:2019ipt}
\be\label{cpglvo0}
\cU_{+,\Delta}^{\msf{a}}(z,\bar{z})=\int_{0}^{\infty}\frac{\d\omega}{\omega^{2-\Delta}}\,\cU_{+}^{\msf{a}}(\kappa=z,\,\tilde{\kappa}=\omega\,\bar{z})\,,
\ee
in the twistor sigma model, with $\cU_+^{\msf{a}}$ the vertex operator for the momentum eigenstate defined by \eqref{glumomeig}. This Mellin transform can be evaluated explicitly to give
\be\label{cgplvo}
\cU_{+,\Delta}^{\msf{a}}(z,\bar{z})=(-\im)^{1-\Delta}\,\Gamma(\Delta-1)\int_{\Sigma} j^{\msf{a}}(\sigma)\,\frac{\delta_{\Delta}(\la\lambda(\sigma)\,z\ra)}{[\mu(\sigma)\,\bar{z}]^{\Delta-1}}\,
\ee
where
\be\label{scaledelt}
\delta_{\Delta}(\la\lambda(\sigma)\,z\ra):=\frac{\la\xi\,\lambda(\sigma)\ra^{\Delta+1}}{\la\xi\,z\ra^{\Delta+1}}\,\delta(\la\lambda(\sigma)\,z\ra)\,,
\ee
for $\xi$ an arbitrary constant spinor arising from performing the scale integral in the twistor wavefunction. 

The vertex operators \eqref{cgplvo} have simple poles for all $\Delta=k\in\Z_{\leq 1}$, corresponding to the infinite tower of conformally soft positive helicity gluons. The corresponding conformally soft gluon vertex operators are defined by taking the residues~\cite{Adamo:2019ipt}:
\be\label{csglvo}
R^{k,\msf{a}}(z,\bar{z}):=\Res{\Delta=k}\,\cU_{+,\Delta}^{\msf{a}}(z,\bar{z})=\frac{\im^{1-k}}{(1-k)!}\int_{\Sigma}j^{\msf{a}}(\sigma)\,[\mu(\sigma)\,\bar{z}]^{1-k}\,\bar{\delta}_{k-1}(\la\lambda(\sigma)\,z\ra)\,.
\ee
As in a trivial background, it is useful to re-label the index $k=3-2p$ so that $2p-2\in\Z_{\geq0}$, and then binomially expand in $[\mu(\sigma)\,\bar{z}]=\mu^{\dot0}(\sigma)+\bar{z}\,\mu^{\dot{1}}(\sigma)$, leaving:
\be\label{csglexp}
R^{3-2p,\msf{a}}(z,\bar{z})=\sum_{m=1-p}^{p-1}\frac{\bar{z}^{p-m-1}\,S^{p,\msf{a}}_{m}(z)}{\Gamma(p-m)\,\Gamma(p+m)}\,,
\ee
where the modes $S^{p,\msf{a}}_{m}(z)$ are defined by
\be\label{sglmode}
S^{p,\msf{a}}_{m}(z)=\int_{\Sigma}J^{\msf{a}}[p+m-1,p-m-1](\sigma)\,\bar{\delta}_{2-2p}(\la\lambda(\sigma)\,z\ra)\,,
\ee
for 
\be\label{Jdef}
J^{\msf{a}}[k,l](\sigma):=\im^{k+l}\,j^{\msf{a}}(\sigma)\,(\mu^{\dot0})^{k}(\sigma)\,(\mu^{\dot1})^{l}(\sigma)\,,
\ee
a basis of $\mathrm{End}\,E[\C^2]$, the $\mathrm{End}\,E$-valued polynomials in $\mu^{\dot\alpha}$.

Now, it follows that the contraction between any two conformally soft positive helicity modes $S^{p,\msf{a}}_{m}(z)$ and $S^{q,\msf{b}}_{n}(z')$ will be governed by the contraction between the $J^{\msf{a}}[k,l]$ currents in the classical twistor sigma model, or the twistor string at level zero~\cite{Adamo:2021zpw}. Using \eqref{scjj} it follows that 
\begin{multline}\label{sglcont1}
S^{p,\msf{a}}_{m}(z)\,S^{q,\msf{b}}_{n}(z')=\int\limits_{\Sigma\times\Sigma'}\frac{\sqrt{\D\sigma\,\D\sigma'}}{(\sigma\,\sigma')}\left(\rho(\sigma)\,\msf{T}^{\msf{a}}\msf{H}(\sigma)\,\msf{H}^{-1}(\sigma')\,\msf{T}^{\msf{b}}\,\bar{\rho}(\sigma')\right. \\
\left.-\rho(\sigma')\,\msf{T}^{\msf{b}}\,\msf{H}^{-1}(\sigma')\,\msf{H}(\sigma)\,\msf{T}^{\msf{a}}\,\bar{\rho}(\sigma)\right)\bar{\delta}_{2-2p}(\la\lambda(\sigma)\,z\ra)\,\bar{\delta}_{2-2q}(\la\lambda(\sigma')\,z'\ra) \\
\times (\mu^{\dot0})^{p+m-1}(\sigma)\,(\mu^{\dot1})^{p-m-1}(\sigma)\,(\mu^{\dot0})^{q+n-1}(\sigma')\,(\mu^{\dot1})^{q-n-1}(\sigma')\,.
\end{multline}
At this point, we can use the trick of \eqref{glholdel} to integrate by parts in $\sigma$ to trade $\bar{\delta}_{2-2p}(\la\lambda(\sigma)\,z\ra)$ for a holomorphic delta function $\bar{\delta}(\sigma\,\sigma')$ against which the integral in $\Sigma'$ can be performed explicitly. 

This fixes $\sigma=\sigma'$ leaving
\begin{multline}\label{sglcont2}
S^{p,\msf{a}}_{m}(z)\,S^{q,\msf{b}}_{n}(z')=-\frac{f^{\msf{abc}}}{\la\xi\,z\ra^{3-2p}}\int_{\Sigma'}J^{\msf{c}}[p+q+m+n-2,p+q-m-n-2](\sigma')\\
\times\,\frac{\la\xi\,\lambda(\sigma')\ra\,\la\xi\,z'\ra^{2-2p}}{\la\lambda(\sigma')\,z\ra}\,\bar{\delta}_{4-2(p+q)}(\la\lambda(\sigma')\,z'\ra) \\
=\frac{f^{\msf{abc}}}{\la z\,z'\ra}\,\frac{\la\xi\,z'\ra^{3-2p}}{\la\xi\,z\ra^{3-2p}}\,S^{p+q-1,\msf{c}}_{m+n}(z')\,,
\end{multline}
with the second line following by using the support of the remaining holomorphic delta function. This relation takes a more familiar form when we work on the affine patch of the celestial sphere and take $\xi^{\alpha}=\iota^{\alpha}=(1,0)$, at which point \eqref{sglcont2} becomes
\be\label{Salg1}
S^{p,\msf{a}}_{m}(z)\,S^{q,\msf{b}}_{n}(z')=\frac{f^{\msf{abc}}}{z-z'}\,S^{p+q-1,\msf{c}}_{m+n}(z')\,.
\ee
This is precisely the Kac-Moody algebra of conformally soft, positive helicity gluons found in a trivial background~\cite{Strominger:2021lvk}. 

The full loop algebra $\cL\mathfrak{g}[\C^2]$ is recovered by expanding the modes $S^{p,\msf{a}}_{m}(z)$ in a formal Laurent series
\be\label{glloop}
S^{p,\msf{a}}_{m}(z)=\sum_{r\in\Z}z^{r-1}\,S^{p,\msf{a}}_{m,r}\,, \qquad S^{p,\msf{a}}_{m,r}:=\frac{1}{2\pi\im}\oint \frac{J^{\msf{a}}[p+m-1,p-m-1](\sigma)}{\la\iota\,\lambda(\sigma)\ra^{2p+r-2}\,\la\lambda(\sigma)\,o\ra^{r}}\,,
\ee
with the contour of integration wrapping $\lambda(\sigma)=z$ and $\la\lambda(\sigma)\,o\ra$ acting as the (complexified) loop parameter. The above calculation then implies that
\be\label{LSalg1}
\left[S^{p,\msf{a}}_{m,r},\,S^{q,\msf{b}}_{n,s}\right]=f^{\msf{abc}}\,S^{p+q-1,\msf{c}}_{m+n,r+s}\,,
\ee
with the bracket $[\cdot,\,\cdot]$ defined by classical Wick contraction in the twistor sigma model. As expected, these are the commutation relations of $\cL\mathrm{End}\,E[\C^2]\cong\cL\mathfrak{g}[\C^2]$, with the isomorphism provided by the holomorphic trivialization of $E$ pulled back to $\Sigma\cong\P^1$.

%%%%%%%%%%%%%%%%%%%%

\subsection{Gravity}

The calculation of the infinite-dimensional chiral algebra for positive helicity gravitons on a SD radiative spacetime proceeds in exactly the same way~\cite{Adamo:2021lrv,Adamo:2021zpw}\footnote{Remarkably, this chiral algebra can also be obtained via double copy of $\cL\mathfrak{g}[\C^2]$ on SD radiative backgrounds~\cite{Brown:2023zxm}.}. In this case, the conformally soft outgoing positive helicity gravitons are represented by vertex operators 
\be\label{csgrvo}
H^{k}(z,\bar{z}):=\frac{\im^{-k}}{(2-k)!}\int_{\Sigma}\frac{\D\sigma}{\mathfrak{s}^2_{\tilde{\mathtt{h}}}(\sigma)}\,[\bmu(\sigma)\,\bar{z}]^{2-k}\,\bar{\delta}_k(\la\lambda(\sigma)\,z\ra)\,,
\ee
for all $k\in\Z_{\leq2}$, obtained from the Mellin transform of the momentum eigenstates \eqref{gravmomeig}. Now we re-label $k=4-2p$ and expand in $\bar{z}$
\be\label{csgrexp}
H^{4-2p}(z,\bar{z})=\sum_{m=1-p}^{p-1}\frac{\bar{z}^{p-m-1}\,w^p_m(z)}{\Gamma(p-m)\,\Gamma(p+m)}\,,
\ee
with the soft graviton modes
\be\label{sgrmode}
w^p_m(z)=\int_{\Sigma}\frac{\D\sigma}{\mathfrak{s}^2_{\tilde{\mathtt{h}}}(\sigma)}\,\mathrm{w}[p+m-1,p-m-1](\sigma)\,\bar{\delta}_{4-2p}(\la\lambda(\sigma)\,z\ra)\,,
\ee
for
\be\label{wdef}
\mathrm{w}[k,l](\sigma):=\im^{k+l-2}\,(\bmu^{\dot0})^{k}(\sigma)\,(\bmu^{\dot1})^{l}(\sigma)\,,
\ee
a basis of polynomials on the $\C^2$ fibres of $\CPT\to\P^1$.

Contractions between the soft graviton modes \eqref{sgrmode} are given by single Wick contractions in the classical twistor sigma model, defined by \eqref{mprop*}. A short calculation shows that
\begin{multline}\label{sgrcont1}
w^p_m(z)\,w^q_n(z')=\int\limits_{\Sigma\times\Sigma'}\frac{\D\sigma\,\D\sigma'}{\mathfrak{s}_{\tilde{\mathtt{h}}}(\sigma)\,\mathfrak{s}_{\tilde{\mathtt{h}}}(\sigma')\,(\sigma\,\sigma')}\, \bar{\delta}_{4-2p}(\la\lambda(\sigma)\,z\ra)\,\bar{\delta}_{4-2q}(\la\lambda(\sigma')\,z'\ra) \\
\times\left[\!\left[\frac{\partial\mathrm{w}}{\partial\mu}[p+m-1,p-m-1](\sigma)\,\frac{\partial\mathrm{w}}{\partial\mu}[q+n-1,q-n-1](\sigma')\right]\!\right]\,,
\end{multline}
and we can now use the trick of \eqref{grholdel} to integrate by parts in $\sigma$ and localize on $\sigma=\sigma'$. In this case,
\begin{multline}
\lim_{\sigma\to\sigma'}\left[\!\left[\frac{\partial\mathrm{w}}{\partial\mu}[p+m-1,p-m-1](\sigma)\,\frac{\partial\mathrm{w}}{\partial\mu}[q+n-1,q-n-1](\sigma')\right]\!\right] \\
=\Big\{\mathrm{w}[p+m-1,p-m-1](\sigma'),\,\mathrm{w}[q+n-1,q-n-1](\sigma')\Big\}\,,
\end{multline}
where
\be\label{poisson}
\Big\{ a,\,b\Big\}=\epsilon^{\dot\alpha\dot\beta}\,\frac{\partial a}{\partial\bmu^{\dot\alpha}}\,\frac{\partial b}{\partial\bmu^{\dot\beta}}\,,
\ee
is the Poisson structure on the fibres of $\CPT\to\P^1$. A straightforward calculation shows that 
\begin{multline}\label{palg}
\Big\{\mathrm{w}[p+m-1,p-m-1](\sigma'),\,\mathrm{w}[q+n-1,q-n-1](\sigma')\Big\} \\
=2\,\Big(m\,(q-1)-n\,(p-1)\Big)\,\mathrm{w}[p+q+m+n-3,p+q-m-n-3](\sigma')\,,
\end{multline}
which allows us to obtain
\begin{multline}\label{sgrcont2}
w^p_m(z)\,w^q_n(z')=-2\,\frac{\big(m\,(q-1)-n\,(p-1)\big)}{\la\xi\,z\ra^{5-2p}}\int_{\Sigma'}\frac{\D\sigma'}{\mathfrak{s}^2_{\tilde{\mathtt{h}}}(\sigma')}\,\frac{\la\xi\,\lambda(\sigma')\ra\,\la\xi\,z'\ra^{4-2p}}{\la\lambda(\sigma')\,z\ra} \\
\times \,\mathrm{w}[p+q+m+n-3,p+q-m-n-3](\sigma') \\
=2\,\frac{\big(m\,(q-1)-n\,(p-1)\big)}{\la z\,z'\ra}\,\frac{\la\xi\,z'\ra^{5-2p}}{\la\xi\,z\ra^{5-2p}}\,w^{p+q-2}_{m+n}(z')\,.
\end{multline}
When evaluated on an affine patch of the celestial sphere with $\xi^{\alpha}=\iota^{\alpha}=(1,0)$, this is revealed to be
\be\label{winf}
w^p_m(z)\,w^q_n(z')=2\,\frac{\big(m\,(q-1)-n\,(p-1)\big)}{z-z'}\,w^{p+q-2}_{m+n}(z')\,,
\ee
defining the Kac-Moody algebra of $w_{1+\infty}$ also found for conformally soft, positive helicity gravitons in Minkowski space~\cite{Strominger:2021lvk}.

This encodes the full loop algebra $\cL\mathfrak{ham}[\C^2]$, as seen by performing the formal Laurent expansion
\be\label{grloop}
w^{p}_{m}(z)=\sum_{r\in\Z}z^{r-1}\,g^{p}_{m,r}\,, \qquad g^p_{m,r}:=\frac{1}{2\pi\im}\oint\frac{\D\sigma}{\mathfrak{s}^2_{\tilde{\mathtt{h}}}(\sigma)}\,\frac{\mathrm{w}[p+m-1,p-m-1](\sigma)}{\la\iota\,\lambda(\sigma)\ra^{2p+r-4}\,\la\lambda(\sigma)\,o\ra^{r}}\,,
\ee
with the calculation above implying 
\be\label{Lhamalg}
\left\{g^p_{m,r},\,g^q_{n,s}\right\}=2\,\big(m\,(q-1)-n\,(p-1)\big)\,g^{p+q-2}_{m+n,r+s}\,.
\ee
This is the loop algebra of the Lie algebra of Hamiltonian vector fields on $\C^2$, generated by the modes $g^{p}_{m,r}$, as desired.

%%%%%%%%%%%%%%%%%%%%
%%%%%%%%%%%%%%%%%%%%

\section{All order holomorphic OPEs in MHV sector} \label{sec6}

We have seen that both the singular holomorphic collinear splitting functions \emph{and} the infinite-dimensional holomorphic chiral algebras of gauge theory and gravity are not deformed in a SD radiative background. This has the important consequence of allowing us to determine celestial OPEs between gluons and gravitons in these backgrounds to \emph{all orders} -- that is, including all regular contributions in the holomorphic collinear limit -- in the MHV sector. The resulting expressions are almost equivalent to those found in~\cite{Adamo:2022wjo,Ren:2023trv} for trivial perturbative backgrounds, although the definition of the soft gluon and graviton descendants (in terms of which the OPE is organized) differ due to the SD radiative background.

\medskip

\paragraph{Yang-Mills:} In the MHV sector, $d=1$ and the twistor sigma model worldsheet is holomorphically identified with the celestial sphere. The twistor sigma model is reduced to an effective 2d CFT on the celestial sphere itself, allowing us to do the worldsheet integrals explicitly~\cite{Adamo:2022wjo}. The effective gluon vertex operators in the MHV sector can be written as:
\be\label{MHVglvo}
\cU_{+}^{\msf{a}}(z,\tilde{\kappa})=j^{\msf{a}}(z)\,\e^{\im\,[\mu(z)\,\tilde{\kappa}]}\,,
\ee
where we abuse notation by using $j^{\msf{a}}$ to denote the conformal weight \emph{zero} pure Kac-Moody current. The single contraction between these currents is given by
\be\label{KMscont}
j^{\msf{a}}(z_i)\,j^{\msf{b}}(z_j)\Big|_{\text{sing. cont.}}=\frac{1}{z_{ij}}\left(\rho_i\,\msf{T}^{\msf{a}}\,\msf{H}_i\,\msf{H}^{-1}_j\,\msf{T}^{\msf{b}}\,\bar{\rho}_j-\rho_j\,\msf{T}^{\msf{b}}\,\msf{H}^{-1}_j\,\msf{H}_i\,\msf{T}^{\msf{a}}\,\bar{\rho}_i\right)\,,
\ee
where $z_{ij}=z_i-z_j$ and $\rho_i\equiv\rho(z_i)$ ($\bar{\rho}_i\equiv\bar{\rho}(z_i)$) is a conformal weight zero fermion in the (anti-)fundamental representation of the gauge group. Note that the expression of the single contraction depends on the explicitly realization of the current algebra. In the leading worldsheet OPE calculation in Section \ref{sec:holOPEgauge}, worldsheet integrals are localized to $z_i\to z_j$, which trivialises the conjugated frames in \eqref{KMscont}, giving the usual Kac-Moody OPE on the right-hand-side. 

However, for contributions regular in $z_{ij}$, a further Taylor expansion is needed. Combining this with regular normal-ordered contributions, the full classical OPE between the Kac-Moody currents is
\begin{multline}
    j^{\msf{a}}(z_i)\,j^{\msf{b}}(z_j)\Big|_{\text{class.}}=\\
    \sum_{n=0}^\infty \frac{z_{ij}^{n-1}}{n!}\left(\partial^n\left(\rho\,\msf{T}^{\msf{a}}\,\msf{H}\right)\msf{H}^{-1}\,\msf{T}^{\msf{b}}\,\bar{\rho}-\partial^n\left(\msf{H}\,\msf{T}^{\msf{a}}\,\bar{\rho}\right)\rho\,\msf{T}^{\msf{b}}\,\msf{H}^{-1} 
   \right) +\sum_{n=0}^{\infty}\frac{z_{ij}^n}{n!}\,:\partial^n j^{\msf{a}}\,j^{\msf{b}}:\,,
\end{multline}
where all remaining fields and functions are implicitly evaluated at $z_j$.

Incorporating this in the contraction between effective vertex operators:
\begin{equation}\label{YM_all_orders_OPE}
    \cU^{\msf{a}}(z_i,\tilde\kappa_i)\,\cU^{\msf{b}}(z_j,\tilde\kappa_j) = \left(\sum_{n=0}^\infty z_{ij}^{n-1}\,\sum_{r=0}^n\frac{1}{r!}\,:j^{\msf{a}}_{r-n}\,j^{\msf{b}}:\,\partial^r\left(\e^{\im[\mu\,i]}\right)\e^{\im[\mu\,j]}  \right)(z_j)\,,
\end{equation}
where we have defined the following Kac-Moody type descendants in the presence of backgrounds:
\begin{equation}
  :j^{\msf{a}}_{-n}\,j^{\msf{b}}: = \begin{cases}
      \rho\,\msf{T}^{\msf{a}}\,\msf{H}\msf{H}^{-1}\,\msf{T}^{\msf{b}}\,\bar{\rho}-\msf{H}\,\msf{T}^{\msf{a}}\,\bar{\rho}\rho\,\msf{T}^{\msf{b}}\,\msf{H}^{-1} & n=0\,;\\
      \frac{1}{(n-1)!}\left(
      \frac{1}{n}\,\partial^n\left(\rho\,\msf{T}^{\msf{a}}\,\msf{H}\right)\msf{H}^{-1}\,\msf{T}^{\msf{b}}\,\bar{\rho}-\frac{1}{n}\,\partial^n\left(\msf{H}\,\msf{T}^{\msf{a}}\,\bar{\rho}\right)\rho\,\msf{T}^{\msf{b}}\,\msf{H}^{-1}+:\partial^{n-1}j^{\msf{a}}j^{\msf{b}}:\right)& n\geq 1\,,
  \end{cases}  
\end{equation}
where we have used $\binom{n-1}{r}\,\frac{1}{(n-1)!}=\frac{1}{r!(n-r-1)!}$. Note that when one sets $z_{ij}\to 0$, the frames drop out and the first case exactly becomes the familiar $f^{\msf{abc}}j^{\msf{c}}$ of the singular holomorphic celestial OPE. In general, although we can write down the full classical OPE between two vertex operators, the regular terms which arise cannot be fully expressed as a new current algebra generator. Indeed, the regular terms depend on the realization of the current algebra rather than the generators themselves.

Nevertheless, \eqref{YM_all_orders_OPE} can be re-expressed in terms of soft gluon descendants. To do this, consider the generators of $\cL\text{End}\,E[\mathbb{C}^2]$, the soft gluon modes $J^{\msf{a}}[q](z,\tilde\kappa)=\frac{\im^q}{q!} j^{\msf{a}}(z)[\mu(z)\tilde\kappa]^q$, and take the classical single contraction with a gluon vertex operator:
\begin{align}
    J^{\msf{a}}[q](z_i,\tilde\kappa_i)\,\cU^{\msf{b}}(z_j,\tilde\kappa_i+\tilde\kappa_j)
    =& \frac{\im^q}{q!}\sum_{n=0}^\infty z_{ij}^{n-1}\,\sum_{p=0}^n\frac{1}{p!}:j^{\msf{a}}_{p-n}\,j^{\msf{b}}:\,\partial^p\left([\mu\,i]^q\right)\e^{\im[\mu\,(i+j))]}  \nonumber\\
    =&\sum_{n=0}^\infty z_{ij}^{n-1}\,J^{\msf{a}}_{-n}[q](\tilde\kappa_i)\,\cU^{\msf{b}}(z_j,\tilde\kappa_i+\tilde\kappa_j)\,,
\end{align}
with
\begin{equation}\label{sgd1}
    J^{\msf{a}}_{-n}[q](\tilde\kappa_i)\,\cU^{\msf{b}}(z_j,\tilde\kappa_i+\tilde\kappa_j) 
    = \frac{\im^q}{q!}\sum_{p=0}^n\frac{1}{p!}\,:j^{\msf{a}}_{p-n}\,j^{\msf{b}}:\,\partial^p\left([\mu\,i]^q\right)\e^{\im[\mu\,(i+j))]}\,,
\end{equation}
defining the soft descendants of the primary gluon.

To match content of \eqref{YM_all_orders_OPE} with \eqref{sgd1}, one can use Hoppe's identity
\begin{equation}\label{Hoppes_identity}
    \partial^n \e^{f(x)} = \e^{f(x)}\,\sum_{l=0}^n\sum_{m=0}^l\frac{(-1)^{l-m}}{(l-m)!m!}\,f(x)^{l-m}\partial^n(f(x)^m)
\end{equation}
on \eqref{YM_all_orders_OPE} and recognize part of it as the soft descendant after identifying $m=q$ and recognizing $r$ and $p$ as the same dummy index. The result (trivially extended to the mixed helicity case) is:
\begin{align}
    \cU_{+}^{\msf{a}}(z_i,\tilde\kappa_i)\,\cU_{\pm}^{\msf{b}}(z_j,\tilde\kappa_j) 
    &= \sum_{n=0}^\infty\sum_{l=0}^n\sum_{m=0}^l\, z_{ij}^{n-1}\,\frac{\left(-\im[\mu\,i]\right)^{l-m}}{(l-m)!}\,J^{\msf{a}}_{-n}[m]\,\cU_{\pm}^{\msf{b}}(z_j,\tilde\kappa_i+\tilde\kappa_j)\nonumber
    \\
    &=\sum_{n=0}^\infty\sum_{l=0}^n\sum_{m=0}^l\, z_{ij}^{n-1}\frac{\left(-[\tilde\kappa_i\frac{\partial}{\partial\tilde\kappa_j}]\right)^{l-m}}{(l-m)!}\,J^{\msf{a}}_{-n}[m]\,\cU_{\pm}^{\msf{b}}(z_j,\tilde\kappa_i+\tilde\kappa_j)\,,
\end{align}
where in the second line object $[\mu\,\tilde\kappa_i]$ has been rewritten as further dotted-momentum derivatives acting on $\cU^{\msf{b}}$. If one were to use an alternative realization of the current algebra, the corresponding definitions of the soft descendant changes accordingly. Overall, one indeed does observe a slight change in the all-orders OPE in the MHV sector due to the presence of the non-trivial background. Additionally, when the SD background is switched off, all holomorphic frames become the identity and the full OPE and definition of the descendants reduces to that given by~\cite{Adamo:2022wjo} in a trivial background.

\medskip

\paragraph{Gravity:}
Similar to the Yang-Mills case, in the presence of generic SD radiative background metric, the positive helicity graviton vertex operator at MHV degree becomes
\begin{equation}
     \mathcal{U}_+(z_i,\tilde\kappa_i)=\frac{1}{\mathfrak{s}^2_{12}(z_i)}
     \,\e^{\im[\bmu(z_i)\,i]}\,,
\end{equation}
where $\bmu^{\dal}=\msf{F}^{\dal}+m^{\dal}$ and $\mathfrak{s}_{12}(z_i)= z_{i1}z_{i2}$. Take a single contraction between two such vertex operators and expand around $z_i=z_j$:
\begin{align}
    \mathcal{U}(z_i,\tilde\kappa_i)\,&\mathcal{U}(z_j,\tilde\kappa_j) = \frac{\tilde\kappa_{i\dal}\tilde\kappa_{j\Dot{\beta}}\,H^{\dal\Dot{\gamma}}(z_i)H^{\Dot{\beta}}_{\Dot{\gamma}}(z_j)}{z_{ij}\,\mathfrak{s}_{12}(z_i)\mathfrak{s}_{12}(z_j)}\, \e^{\im[\bmu(z_i)\,i]}\,\e^{\im[\bmu(z_j)\,j]}\\
    & =\sum_{n=0}^{\infty} \frac{z_{ij}^{n-1}}{n!}\,\frac{\tilde\kappa_{i\dal}\,\tilde\kappa_{j\Dot{\beta}}\,H^{\Dot{\beta}}_{\Dot{\gamma}}(z_j)}{\mathfrak{s}_{12}(z_j)}\,\partial^n\left(\frac{\e^{\im[\bmu\,i]}H^{\dal\Dot{\gamma}}}{\mathfrak{s}_{12}}\right)\e^{\im[\bmu(z_j)\,j]}\nonumber\\
    & = \sum_{n=0}^{\infty}\sum_{p=0}^n\binom{n}{p} \frac{z_{ij}^{n-1}}{n!\,\mathfrak{s}_{12}(z_j)} \,\tilde\kappa_{i\dal}\,\tilde\kappa_{j\Dot{\beta}}\,H^{\Dot{\beta}}_{\Dot{\gamma}}\,\partial^{n-p}\left(\e^{\im[\bmu\,i]}\right)\partial^{p}\left(\frac{H^{\dal\Dot{\gamma}}}{\mathfrak{s}_{12}}\right)\,\e^{\im[\bmu\,j]}\,,\nonumber
\end{align}
with all functions evaluated at $z_j$ in the final line. Notice that holomorphic frames evaluated at $z_i$ are also Laurent expanded. Further expanding the factor $\partial_{z_i}^p\left(\e^{\im[\bmu\,i]}\right)$ using \eqref{Hoppes_identity}, we have:
\begin{multline}\label{bkgd_full_OPE}
    \cU(z_i,\tilde\kappa_i)\,\cU(z_j,\tilde\kappa_j) = \sum_{n=0}^{\infty}\sum_{p=0}^n\sum_{l=0}^p\sum_{m=0}^l\binom{n}{p} \frac{z_{ij}^{n-1}}{n!\,\mathfrak{s}_{12}(z_j)}\,\tilde\kappa_{i\dal}\tilde\kappa_{j\Dot{\beta}}\,H^{\Dot{\beta}}_{\Dot{\gamma}}\,\frac{(-1)^{l-m}\im^l}{(l-m)!m!}\\
    \times\,\left([\bmu\,i]^{l-m}\right)\partial^{n-p}\left([\bmu\,i]^m\right)\partial^{p}\left(\frac{H^{\dal\Dot{\gamma}}}{\mathfrak{s}_{12}}\right)\e^{\im[\bmu(i+j)]}\,.
\end{multline}
In order to rewrite this full classical OPE as descendants of $\mathcal{U}$, we consider the following soft operator
\begin{equation}
    W[q](z,\tilde\kappa) = \frac{1}{\mathfrak{s}^2_{12}(z)}\frac{\im^q}{q!} [\bmu(z)\tilde\kappa]^q \,,
\end{equation}
and its single contraction with $\mathcal{U}(z_j,\tilde\kappa_i+\tilde\kappa_j)$:
\begin{align}
    &W[q](z_i,\tilde\kappa_i)\,\cU(z_j,\tilde\kappa_i+\tilde\kappa_j) \nonumber\\
    & = \sum_{n=0}^{\infty}\sum_{r=0}^n\binom{n}{r} \frac{z_{ij}^{n-1}}{n!\,\mathfrak{s}_{12}(z_j)}\,\tilde\kappa_{i\dal}\tilde\kappa_{j\Dot{\beta}}\,H^{\Dot{\beta}}_{\Dot{\gamma}}\,\frac{\im^q}{(q-1)!}\partial_{z_i}^{n-r}\left([\bmu\,i]^{q-1}\right)\partial^{r}\left(\frac{H^{\dal\Dot{\gamma}}}{\mathfrak{s}_{12}}\right)\e^{\im[\bmu(i+j)]}\nonumber\\
    &=\sum_{n=0}^{\infty}\sum_{r=0}^n\binom{n}{r} \frac{z_{ij}^{n-1}}{n!\,\mathfrak{s}_{12}(z_j)}\,\tilde\kappa_{j\Dot{\beta}}\,H^{\Dot{\beta}}_{\Dot{\gamma}}\,\widetilde W_{-n,r}^{\Dot{\gamma}}[q]\,\mathcal{U}(z_j,\tilde\kappa_i+\tilde\kappa_j)\,,
\end{align}
where we have defined the following soft graviton descendant in a SD radiative spacetime:
\begin{equation}\label{bkgd_gravity_descendant}
    \widetilde W_{-n,r}^{\Dot{\gamma}}[q]\,\mathcal{U}(z_j,\tilde\kappa_i+\tilde\kappa_j) =\frac{\im^q\,\tilde\kappa_{i\dal}}{(q-1)!}\partial^{n-r}\left([\bmu\,i]^{q-1}\right)\partial^r\left(\frac{H^{\dal\Dot{\gamma}}}{\mathfrak{s}_{12}}\right)\e^{\im[\bmu(i+j)]}\,.
\end{equation}
Comparing \eqref{bkgd_gravity_descendant} with terms in the full OPE \eqref{bkgd_full_OPE}, we see that we just need to set $q=m+1$ and $r=p$ to rewrite \eqref{bkgd_full_OPE} as:
\begin{align}
    &\mathcal{U}_{+}(z_i,\tilde\kappa_i)\,\mathcal{U}_{\pm}(z_j,\tilde\kappa_j) \label{gallorder_final} \\
    =& \sum_{n=0}^{\infty}\sum_{p=0}^n\sum_{l=0}^p\sum_{m=0}^l\binom{n}{p} \frac{z_{ij}^{n-1}}{n!\,\mathfrak{s}_{12}(z_j)}\,\tilde\kappa_{j\Dot{\beta}}\,H^{\Dot{\beta}}_{\Dot{\gamma}}\,\frac{(-\im)^{l-m+1}}{(l-m)!}\,[\bmu\,i]^{l-m}\,\widetilde W^{\Dot{\gamma}}_{-n,p}[m+1]\,\mathcal{U}_{\pm}(z_j,\tilde\kappa_i+\tilde\kappa_j)\nonumber\\
    =& \sum_{n=0}^{\infty}\sum_{p=0}^n\sum_{l=0}^p\sum_{m=0}^l\binom{n}{p} \frac{z_{ij}^{n-1}}{n!\,\mathfrak{s}_{12}(z_j)}\,\tilde\kappa_{j\Dot{\beta}}\,H^{\Dot{\beta}}_{\Dot{\gamma}}\,\frac{-\im\left(-[\tilde\kappa_i\frac{\partial}{\partial\tilde\kappa_j}]\right)^{l-m}}{(l-m)!}\,\widetilde W^{\Dot{\gamma}}_{-n,p}[m+1]\,\mathcal{U}_{\pm}(z_j,\tilde\kappa_i+\tilde\kappa_j)\nonumber\,,
\end{align}
including the trivial extension to mixed-helicity. In the second equality, polynomials in $[\bmu\tilde\kappa_i]$ have been traded for derivatives acting on the soft descendant of $\mathcal{U}$. Turning off the background sets all the frames $H^{\dot\alpha}{}_{\dot\beta}=\delta^{\dot\alpha}_{\dot\beta}$, and after a making an alternative gauge choice for the graviton vertex operators (which is allowed in Minkowski spacetime) to set $\mathfrak{s}_{12}(z_i)=1$, \eqref{gallorder_final} gives the full graviton OPE in flat space derived using BCFW recursions in~\cite{Ren:2023trv}.

%%%%%%%%%%%%%%%%%%%%%
%%%%%%%%%%%%%%%%%%%%%

\section{Discussion}
\label{sec:dis}

In this paper, we studied the holomorphic infrared structures of gluon and graviton scattering amplitudes in self-dual radiative backgrounds. To do this we used two distinct approaches: explicit, all-multiplicity amplitude formulae for the tree-level MHV sector, and the semiclassical twistor sigma models which underpin them. Remarkably, we found that holomorphic collinear splitting functions (and consequently, holomorphic celestial OPE coefficients) and infinite-dimensional chiral algebras associated to the conformally soft SD sectors are un-deformed by the backgrounds. Within the MHV sector, the all-order holomorphic celestial OPE, containing all regular contributions to the collinear limit, \emph{is} deformed by the background, but remains explicitly calculable in terms of soft descendants.

From the perspective of background field theory, these results are remarkable: SD radiative backgrounds are specified by unconstrained functional degrees of freedom. However, from the perspective of twistor theory they are perhaps less surprising: the twistorial complex structures associated to SD radiative backgrounds have global representations in terms of the radiative data. Motivated by our findings, one can make the following conjecture:
\begin{conjecture}
The holomorphic IR structures of perturbative gauge theory and gravity only differ between choices of SD background when the backgrounds differ by the presence of a source or large data. Equivalently, they differ only if the associated twistor descriptions are not related by a globally well-defined Dolbeault representative on twistor space.
\end{conjecture}
For instance, in the context of general relativity, our work establishes that the conjecture is true for Minkowski spacetime and any SD radiative spacetime.  It would be interesting to explore strategies to prove this conjecture more generally.

It is worth noting that in the cases considered in this paper, the conjecture aligns nicely with basic physical intuition. Indeed, in the perturbative limit, a SD radiative background is simply a single gluon or graviton of positive helicity propagating in an otherwise trivial configuration. Thus, in this perturbative regime, it is unsurprising that the holomorphic IR structure are un-modified. For SD plane waves, which are coherent superpositions of such positive helicity gluons or gravitons, this intuition is fairly precise. 

\medskip

There are many other interesting directions to explore, following on from this work. It is known that for backgrounds which break translation invariance, the associated low-multiplicity celestial amplitudes are less singular than those in a trivial background, and often take a simple form~\cite{Fan:2022vbz,Casali:2022fro,Fan:2022kpp,deGioia:2022fcn,Gonzo:2022tjm,Stieberger:2022zyk,Melton:2022fsf,Banerjee:2023rni,Taylor:2023bzj,Stieberger:2023fju}. A generic SD radiative background also breaks translation invariance, and in special cases (e.g., a SD plane wave) it may be possible to evaluate the celestial amplitude explicitly to see if they also have improved analytic properties.

This could have some interesting consequences. For instance, scattering of gluons with leading-order Faddeev-Kulish dressing~\cite{Chung:1965zza,Kibble:1968oug,Kibble:1968npb,Kibble:1968lka,Kulish:1970ut,Kapec:2017tkm} in a trivial background can be thought of as scattering gluons in certain QED backgrounds. An interesting relation exists between such dressing and the holomorphic frames furnishing states on twistor space~\cite{Bu:2023hhh}. This gives an alternative way to interpret the generic SD radiative backgrounds we study as some Faddeev-Kulish-type dressing of gluon vertex operators. If the associated celestial amplitudes have improved analytic properties, then it suggests a connection between resolutions of IR singularities and singularities of celestial amplitudes in the conformal primary basis.

It is also important to explore the fate of IR structures for \emph{non-chiral} backgrounds. A critical obstruction to this is that the twistor descriptions used here do not extend to non-trivial, Lorentzian-real backgrounds at the non-linear level. Thus, to proceed one requires explicit scattering amplitude formulae, determined by the background-field Feynman rules, at 4-points (at least). For pure Yang-Mills theory in a non-chiral plane wave gauge field background, this computation has been done~\cite{Adamo:2018mpq}, meaning that the raw tools to undertake such a study are already in place.

Finally, it would be interesting to furnish a derivation, either from first principles or twistor string theory, of the generalized graviton scattering formula for SD radiative spacetimes presented in Appendix~\ref{appA}, and used in Section~\ref{sec:holOPEgravity}. On the twistor theory side, this requires a better understanding of how to couple the twistor string for gravity~\cite{Skinner:2013xp} to strong SD background fields.

\acknowledgments

We thank Simon Heuveline, Atul Sharma and David Skinner for interesting conversations. TA is supported by a Royal Society University Research Fellowship, the  Leverhulme Trust grant RPG-2020-386 and the Simons Collaboration on Celestial Holography MP-SCMPS-00001550-11. WB is supported by a Royal Society PhD Studentship. BZ is supported by the Royal Society.

%%%%%%%%%%%%%%%%%%%%%
%%%%%%%%%%%%%%%%%%%%%

\appendix

\section{Generalised formula for graviton scattering} \label{appA}

The formulae given in~\cite{Adamo:2022mev} for tree-level graviton scattering amplitudes on a SD radiative spacetime has a rather restrictive form when compared with the analogous formulae in Minkowski spacetime~\cite{Hodges:2011wm,Hodges:2012ym,Cachazo:2012kg}. In particular, the diagonal entries of the matrix $\cH$ require a choice of $d+1$ reference points on $\P^1$, which in~\cite{Adamo:2022mev} are restricted to be the those associated with the negative helicity gravitons; this is a technical consequence of the derivation of these formulae rooted in hyperk\"ahler geometry~\cite{Adamo:2021bej}. By contrast, in Minkowski space these reference points can be freely chosen as a consequence of momentum conservation.

Even in Minkowski space, this level of generality is important for manifesting generic collinear or soft limits (cf., \cite{Bullimore:2012cn,Cachazo:2012pz}). However, in generic SD radiative spacetimes there is no overall momentum conservation, so it is not immediately clear how to generalise the formulae of~\cite{Adamo:2022mev} to having arbitrary reference spinors. In this appendix, we present a conjecture for such a generalisation: while we do not have any sort of first-principle derivation of the result, it passes the basic test of reducing to the known formulae when the reference spinors coincide with the locations of the negative helicity gravitons, and gives the correct holomorphic collinear splitting functions. A proof -- or correction -- of this formula requires a proper understanding of the gravitational twistor string~\cite{Skinner:2013xp} in a SD radiative background.

\medskip

Our approach is motivated by the analogous extension of the Minkowski space gravitational S-matrix to (A)dS~\cite{Adamo:2015ina}: that is, by systematically correcting the amplitude formula to ensure that reference points can be chosen arbitrarily. To being, let us consider the na\"ive extension of the N$^{d-1}$MHV amplitude formula on a SD radiative spacetime with arbitrary reference points (trivially extended to $\cN=8$ supergravity for simplicity):
\begin{multline}\label{mod1}
C^{(0)}_{n,d}=\sum_{t,\{p_t\}}\int\frac{\d^{4|8(d+1)}U}{\mathrm{vol\,\GL(2,\C)}}\,\mathrm{det}'\!\left(\HH^{\vee}\right)\left(\prod_{\m=1}^{t}\,\frac{\partial^{p_\m}}{\partial\varepsilon_\m^{p_\m}}\right)\mathrm{det}^{\prime}(\cG)\Big|_{\varepsilon=0}\\
\times \prod_{i=1}^{n}h_{i}(Z(\sigma_i))\,\prod_{\m=1}^{t}\D\bar{\lambda}(\sigma_m)\,N^{(p_\m-2)}(\sigma_\m)\,,
\end{multline}
where the distinction with the previous formula is in the definition of
\be\label{moddetprime}
\mathrm{det}^{\prime}(\cG):=\frac{\left|\cG^{\tilde{\mathtt{h}}\cup k}_{\tilde{\mathtt{h}}\cup k}\right|}{|\tilde{\mathtt{h}}\cup k|^2}\,\prod_{j\in\{\tilde{\mathtt{h}}\cup k\}}\D\sigma_j\,,
\ee
where $\tilde{\mathtt{h}}$ is the set of $d+1$ negative helicity gravitons and $k$ is an arbitrarily chosen positive helicity graviton. Matrix $\cG$ has the same block decomposition as $\cH$ \eqref{cHblock}, but the diagonal entries are now defined using a set of, in principle arbitrary, $d+1$ reference points $\{\sigma_{s_0},\ldots,\sigma_{s_d}\}$ on $\P^1$:
\be\label{modHH}
\HH_{ii}:=t_{i}\,\D\sigma_i\sum_{j\neq i}\,\frac{t_j\,[\![i\,j]\!]}{(i\,j)}\prod_{r=0}^{d}\,\frac{(s_r\,j)}{(s_r\,i)}-\im\,t_i\,\D\sigma_i\sum_{\m=1}^{t}\,\frac{\varepsilon_\m\,[\![i\,\bar{\lambda}(\sigma_\m)]\!]}{(i\,\m)}\prod_{r=0}^{d}\,\frac{(s_r\,\m)}{(s_r\,i)}\,,
\ee
\be\label{modTT}
\T_{\m\m}:=-\im\,\varepsilon_\m\,\D\sigma_\m\sum_{i=1}^{n}\,\frac{t_i\,[\![\bar{\lambda}(\sigma_\m)\,i]\!]}{(\m\,i)}\,\prod_{r=0}^{d}\,\frac{(s_r\,i)}{(s_r\,\m)}-\varepsilon_\m\,\D\sigma_\m\sum_{\mathrm{l}\neq\m}\,\frac{\varepsilon_{\mathrm{l}}\,[\![\bar{\lambda}(\sigma_\m)\,\bar{\lambda}(\sigma_{\mathrm{l}})]\!]}{(\m\,\mathrm{l})}\,\prod_{r=0}^{d}\,\frac{(s_r\,\mathrm{l})}{(s_r\,\m)}\,.
\ee
The choice of these reference points has no physical meaning (the scattering amplitudes cannot depend on them), so in order for \eqref{mod1} to make sense it must not depend on them.

This can be checked by differentiating $C^{(0)}_{n,d}$ with respect to any one of the reference spinors. Without loss of generality, let $\xi^{\ba}\equiv\sigma^{\ba}_{s_0}$ and $\d_{\xi}:=\d\xi^{\ba}\,\frac{\partial}{\partial\xi^{\ba}}$ denote the exterior differential. A straightforward calculation using Jacobi's formula reveals that
\begin{multline}\label{dmod1}
\d_{\xi}C^{(0)}_{n,d}=\sum_{t,\{p_t\}}\int\frac{\d^{4|8(d+1)}U}{\mathrm{vol\,\GL(2,\C)}}\,\mathrm{det}'\!\left(\HH^{\vee}\right)\left(\prod_{\m=1}^{t}\,\frac{\partial^{p_\m}}{\partial\varepsilon_\m^{p_\m}}\right)\left[\sum_i\mathrm{det}'(\cG^{i}_{i})\,\d_{\xi}\HH_{ii}\right. \\
\left.\left.+\sum_{\m=1}^{t}\mathrm{det}'(\cG^{\m}_{\m})\,\d_{\xi}\T_{\m\m}\right]\right|_{\varepsilon=0}\prod_{i=1}^{n}h_{i}(Z(\sigma_i))\,\prod_{\m=1}^{t}\D\bar{\lambda}(\sigma_m)\,N^{(p_\m-2)}(\sigma_\m)\,,
\end{multline}
where
\be\label{dHH}
\d_{\xi}\HH_{ii}=-\frac{t_i\,\D\sigma_i\,\D\xi}{(\xi\,i)^2}\left[\sum_{j=1}^{n}t_j\,[\![i\,j]\!]\,\prod_{r=1}^{d}\,\frac{(s_r\,j)}{(s_r\,i)}-\sum_{\m=1}^{t}\varepsilon_\m\,[\![i\,\bar{\lambda}(\sigma_\m)]\!]\,\prod_{r=1}^{d}\,\frac{(s_r\,\m)}{(s_r\,i)}\right]\,,
\ee
and
\be\label{dTT}
\d_{\xi}\T_{\m\m}=\frac{\varepsilon_\m\,\D\sigma_\m\,\D\xi}{(\xi\,\m)^2}\left[\im\sum_{j=1}^{n}t_j\,[\![\bar{\lambda}(\sigma_\m)\,j]\!]\,\prod_{r=1}^{d}\,\frac{(s_r\,j)}{(s_r\,\m)}+\sum_{\mathrm{l}=1}^{t}\varepsilon_{\mathrm{l}}\,[\![\bar{\lambda}(\sigma_{\m})\,\bar{\lambda}(\sigma_{\mathrm{l}})]\!]\,\prod_{r=1}^{d}\,\frac{(s_r\,\mathrm{l})}{(s_r\,\m)}\right]\,,
\ee
after applying the Schouten identity. By exploiting the fact that
\be\label{diffmod}
\frac{\partial\mu^{\dot\alpha}(\sigma)}{\partial U^{\dot\beta\ba(d)}}=H^{\dot\alpha}{}_{\dot\beta}(U,\sigma)\,\sigma_{\ba(d)}\,,
\ee
the terms in \eqref{dmod1} can be rewritten as moduli derivatives acting on the graviton wavefunctions and news functions:
\begin{multline}\label{dmod1*}
\d_{\xi}C^{(0)}_{n,d}=\D\xi\sum_{t,\{p_t\}}\int\frac{\d^{4|8(d+1)}U}{\mathrm{vol\,\GL(2,\C)}}\,\mathrm{det}'\!\left(\HH^{\vee}\right)\left(\prod_{\m=1}^{t}\,\frac{\partial^{p_\m}}{\partial\varepsilon_\m^{p_\m}}\right) \\
\left.\left[\im\sum_i \mathrm{det}'(\cG^{i}_{i})\,\frac{t_i\,\D\sigma_i}{(\xi\,i)^2}\,\frac{\tilde{K}^{\dot\alpha}_{i}\,\sigma_{s_1}^{\ba_1}\cdots\sigma_{s_d}^{\ba_d}}{(s_1\,i)\cdots(s_d\,i)}+\sum_{\m=1}^{t}\mathrm{det}'(\cG^{\m}_{\m})\,\frac{\varepsilon_\m\,\D\sigma_\m}{(\xi\,\m)^2}\,\frac{\bar{\Lambda}^{\dot\alpha}_\m\,\sigma_{s_1}^{\ba_1}\cdots\sigma_{s_d}^{\ba_d}}{(s_1\,\m)\cdots(s_d\,\m)}\right]\right|_{\varepsilon=0} \\
\times\,\frac{\partial}{\partial U^{\dot\alpha\ba(d)}}\prod_{i=1}^{n}h_{i}(Z(\sigma_i))\,\prod_{\m=1}^{t}\D\bar{\lambda}(\sigma_m)\,N^{(p_\m-2)}(\sigma_\m)\,,
\end{multline}
where we have employed the shorthand
\be\label{dmomsh}
\tilde{K}^{\dot\alpha}_{i}:=\tilde{\kappa}_{i\,\dot\beta}\,H^{\dot\beta\dot\alpha}(U,\sigma_i)\,, \qquad \bar{\Lambda}^{\dot\alpha}_\m:=\bar{\lambda}_{\dot\beta}(\sigma_\m)\,H^{\dot\beta\dot\alpha}(U,\sigma_\m)\,,
\ee
to denote background dressed dotted spinors.

In Minkowski spacetime (where all tail terms in this expression vanish and the background dressings become trivial), this would imply that the whole integrand of \eqref{dmod1*} is a total derivative with respect to the map moduli, and thus that the variation of the formula with respect to any of the reference spinors vanishes. However, for a SD radiative spacetime, we cannot simply pull the moduli derivative past the middle line of \eqref{dmod1*}, where each term contains complicated moduli dependence through the background dressing of all entries in the matrix $\cH$. In other words, $\d_{\xi}C^{(0)}_{n,d}\neq 0$, so is not independent of the choice of reference spinors. To rectify this, we must add further correction terms to \eqref{mod1}.

Define
\begin{multline}\label{corrterm}
C^{(\gamma)}_{n,d}:=\frac{(-1)^\gamma}{\gamma!}\,\sum_{t,\{p_t\}}\int\frac{\d^{4|8(d+1)}U}{\mathrm{vol\,\GL(2,\C)}}\,\mathrm{det}'\!\left(\HH^{\vee}\right)\left(\prod_{\m=1}^{t}\,\frac{\partial^{p_\m}}{\partial\varepsilon_\m^{p_\m}}\right) \\
\sum_{b=0}^{\gamma} \sum_{\substack{i_1,\ldots,i_b \\ \m_1,\ldots,m_{\gamma-b}}}f_{i_1}^{\ba_1(d)}\cdots f_{i_b}^{\ba_b(d)}\,f_{\m_1}^{\ba_{b+1}(d)}\cdots f_{\m_{\gamma-b}}^{\ba_d(d)}\\
\frac{\partial}{\partial U^{\dot\alpha_1\ba_1(d)}}\cdots\frac{\partial}{\partial U^{\dot\alpha_\gamma\ba_\gamma(d)}}\left(\tilde{K}_{i_1}^{\dot\alpha_1}\cdots\tilde{K}_{i_b}^{\dot\alpha_b}\,\bar{\Lambda}^{\dot\alpha_{b+1}}_{\m_1}\cdots\bar{\Lambda}^{\dot\alpha_\gamma}_{\m_{\gamma-b}}\,\mathrm{det}'(\cG^{i_1\cdots i_{b}\m_1\cdots\m_{\gamma-b}}_{i_1\cdots i_{b}\m_1\cdots\m_{\gamma-b}})\right)\Big|_{\varepsilon=0}\\
 \prod_{i=1}^{n}h_{i}(Z(\sigma_i))\,\prod_{\m=1}^{t}\D\bar{\lambda}(\sigma_m)\,N^{(p_\m-2)}(\sigma_\m)\,,
\end{multline}
where
\be\label{fdef}
f^{\ba(d)}_{i}:=\im\,t_i\,\frac{\D\sigma_i}{d+1}\sum_{a\in\mathtt{g}}\sum_{r=0}^{d}\,\frac{(s_r\,a)}{(s_r\,i)\,(a\,i)}\,\prod_{p\neq r}\,\frac{\sigma_{s_p}^{\ba_p}}{(s_p\,i)}\,,
\ee
for $\mathtt{g}$ any set of $d+1$ of the external gravitons, and $f_{\m}^{\ba(d)}$ is given by simply replacing $\im t_i\to\varepsilon_\m$ and $i\to\m$ everywhere else in this expression. The utility of these definitions is revealed by observing that
\begin{multline}\label{df}
\d_{\xi}f_{i}^{\ba(d)}=\im\,t_i\,\frac{\D\sigma_i\,\D\xi}{(d+1)\,(\xi\,i)^2}\sum_{a\in\mathtt{g}}\left[\sum_{r=1}^{d}\,\frac{(s_r\,a)\,\sigma_i^{(\ba}}{(s_r\,i)\,(a\,i)}\,\prod_{p\neq0,r}\,\frac{\sigma_{s_p}^{\ba_{s_p})}}{(s_p\,i)}-\prod_{p=1}^{d}\,\frac{\sigma_{s_p}^{\ba_p}}{(s_p\,i)}\right] \\
=-\im\,t_{i}\,\D\sigma_i\,\D\xi\prod_{p=1}^{d}\,\frac{\sigma_{s_p}^{\ba_p}}{(s_p\,i)}+\im\,t_i\,\frac{\D\sigma_i\,\D\xi}{(d+1)\,(\xi\,i)^2}\sum_{a\in\mathtt{g}}\sum_{r=1}^{d}\,\frac{(s_r\,a)\,\sigma_i^{(\ba}}{(s_r\,i)\,(a\,i)}\,\prod_{p\neq0,r}\,\frac{\sigma_{s_p}^{\ba_{s_p})}}{(s_p\,i)}\,.
\end{multline}
In \eqref{corrterm}, $f_{i}^{\ba(d)}$ is always contracted via a moduli derivative $\frac{\partial}{\partial U^{\dot\alpha\ba(d)}}$ with the dressed momentum spinor $\tilde{K}^{\dot\alpha}_{i}$, which means that contributions from \eqref{df} proportional to $\sigma_i^{\ba}$ will be projected to zero. So, effectively we have that
\be\label{df*}
\d_{\xi}f_{i}^{\ba(d)}\simeq-\im\,t_{i}\,\D\sigma_i\,\D\xi\prod_{p=1}^{d}\,\frac{\sigma_{s_p}^{\ba_p}}{(s_p\,i)}\,,
\ee
with `$\simeq$' denoting equal up to terms that vanish inside of the full formula \eqref{corrterm}. Similar formulae hold in the obvious way for $f_\m^{\ba(d)}$.

Using this, it follows that
\begin{multline}\label{dC1}
\d_{\xi}C^{(1)}_{n,d}=\D\xi\sum_{t,\{p_t\}}\int\frac{\d^{4|8(d+1)}U}{\mathrm{vol\,\GL(2,\C)}}\,\mathrm{det}'\!\left(\HH^{\vee}\right)\left(\prod_{\m=1}^{t}\,\frac{\partial^{p_\m}}{\partial\varepsilon_\m^{p_\m}}\right) \\
\frac{\partial}{\partial U^{\dot\alpha\ba(d)}}\left[\im\sum_i \mathrm{det}'(\cG^{i}_{i})\,\frac{t_i\,\D\sigma_i}{(\xi\,i)^2}\,\frac{\tilde{K}^{\dot\alpha}_{i}\,\sigma_{s_1}^{\ba_1}\cdots\sigma_{s_d}^{\ba_d}}{(s_1\,i)\cdots(s_d\,i)}\right. \\
\left.\left.+\sum_{\m=1}^{t}\mathrm{det}'(\cG^{\m}_{\m})\,\frac{\varepsilon_\m\,\D\sigma_\m}{(\xi\,\m)^2}\,\frac{\bar{\Lambda}^{\dot\alpha}_\m\,\sigma_{s_1}^{\ba_1}\cdots\sigma_{s_d}^{\ba_d}}{(s_1\,\m)\cdots(s_d\,\m)}\right]\right|_{\varepsilon=0}\prod_{i=1}^{n}h_{i}(Z(\sigma_i))\,\prod_{\m=1}^{t}\D\bar{\lambda}(\sigma_m)\,N^{(p_\m-2)}(\sigma_\m) \\
-\sum_{t,\{p_t\}}\int\frac{\d^{4|8(d+1)}U}{\mathrm{vol\,\GL(2,\C)}}\,\mathrm{det}'\!\left(\HH^{\vee}\right)\left(\prod_{\m=1}^{t}\,\frac{\partial^{p_\m}}{\partial\varepsilon_\m^{p_\m}}\right) \left[\sum_{i}f_{i}^{\ba(d)}\,\frac{\partial}{\partial U^{\dot\alpha\ba(d)}}\d_{\xi}\left(\tilde{K}_{i}^{\dot\alpha}\mathrm{det}'(\cG^{i}_{i})\right)\right.\\
\left.\left.+\sum_{\m=1}^t f_{\m}^{\ba(d)}\,\frac{\partial}{\partial U^{\dot\alpha\ba(d)}}\d_{\xi}\left(\bar{\Lambda}_{\m}^{\dot\alpha}\,\mathrm{det}'(\cG^{\m}_{\m})\right)\right]\right|_{\varepsilon=0}\prod_{i=1}^{n}h_{i}(Z(\sigma_i))\,\prod_{\m=1}^{t}\D\bar{\lambda}(\sigma_m)\,N^{(p_\m-2)}(\sigma_\m)\,.
\end{multline}
The first three lines of this expression combine with $\d_{\xi}C^{(0)}_{n,d}$ from \eqref{dmod1*} to give a contribution which vanishes as a total derivative on the map moduli space. The final two lines give terms which are themselves completed to total derivatives by contributions coming from $\d_{\xi}C^{(2)}_{n,d}$. In this way, one proceeds inductively to see that dependence on the reference spinor is cancelled between successive terms in the series of $C^{(\gamma)}_{n,d}$ corrections. This iterative cancellation terminates when the reduced determinant is exhausted.

This means that the expression
\be\label{newamp}
\boxed{\cM_{n,d}=\sum_{\gamma=0}^{2(n-d)-5}C^{(\gamma)}_{n,d}\,,}
\ee
obeys $\d_{\xi}\cM_{n,d}=0$, with the range of the sum dictated by the maximum size of the matrix $\cH$ for fixed $n$ and $d$. This in turn implies that in $\cM_{n,d}$, the choice of reference points $\{\sigma_{s_r}\}$ is entirely arbitrary. In other words, we have constructed a candidate formula for the tree-level N$^{d-1}$MHV graviton scattering amplitude in a SD radiative spacetime which has arbitrarily-chosen reference points. However, it is important to note that this construction is certainly \emph{not} a derivation: we proceeded simply by making a na\"ive modification to the known formula and then systematically correcting it to obtain an expression which is independent of the choice of reference points. In this sense, \eqref{newamp} is only a conjecture.

Nevertheless, it can be shown that \eqref{newamp} passes several basic consistency tests required of such a candidate amplitude formula. The first of these is the flat space limit, where the background is Minkowski spacetime. In this case, the $C^{(\gamma\geq1)}_{n,d}=0$, as the moduli derivatives in \eqref{corrterm} act on an expression which no longer has any moduli dependence. Indeed, in Minkowski space $H^{\dot\alpha}_{\dot\beta}\to\delta^{\dot\alpha}_{\dot\beta}$ so all moduli dependence drops out of the matrix $\cG$, and $\cM_{n,d}$ reduces to $C^{(0)}_{n,d}$ evalutaed for flat space. It is straightforward to see that this is in fact equal to the Cachazo-Skinner formula~\cite{Cachazo:2012kg} for the tree-level graviton S-matrix in Minkowski space (written with arbitrary reference spinors).

A second non-trivial test passed by \eqref{newamp} is that it reduces to the formula of~\cite{Adamo:2022mev} when the reference points on $\P^1$ are chosen to coincide with the negative helicity gravitons. To see this, one simply takes the arbitrary set of $d+1$ gravitons, $\mathtt{g}$, in \eqref{fdef} to be equal to the set of negative helicity gravitons, $\tilde{\mathtt{h}}$. Then taking the reference points to coincide with the negative helicity gravitons as well, $\{\sigma_{s_r}\}=\tilde{\mathtt{h}}$, gives
\be\label{freduce}
f_{i}^{\ba(d)}=\im\,t_i\,\frac{\D\sigma_i}{d+1}\sum_{a,b\in\tilde{\mathtt{h}}}\,\frac{(b\,a)}{(b\,i)\,(a\,i)}\,\prod_{p\neq b}\,\frac{\sigma_{p}^{\ba_p}}{(p\,i)}=0\,,
\ee
which vanishes as the sum is symmetric in $a,b\in\tilde{\mathtt{h}}$ but the summands are skew. So, with these choices $C^{(\gamma\geq1)}_{n,d}=0$ and \eqref{newamp} immediately reduces to the previously known formula. 
 
\newpage
\bibliographystyle{JHEP}
\bibliography{cope.bib}

\end{document}